\documentclass[12pt]{article}

\usepackage[cp1251]{inputenc}
\usepackage[T2A]{fontenc}
\usepackage[english]{babel}
\usepackage{amsmath,amsfonts,amssymb}
\usepackage{amsbsy}
\usepackage{bm}

\usepackage{graphicx}
\usepackage{cite}
\usepackage{epstopdf}

\newtheorem{thm}{Theorem}

\newtheorem{Def}{Definition}

\renewcommand{\vec}[1]{\boldsymbol{#1}}

\newcommand {\diag} {\mathop {\rm diag} \nolimits}

\newcommand {\const} {\mathop {\rm const} \nolimits}

\begin{document}

\begin{center}
\Large\textsc{\textbf{Existence of Liouvillian solutions in the Hess - Sretensky case of the problem of motion of a gyrostat with a fixed point.}}
\end{center}
{\vskip 0.5cm}
\begin{center}
{\Large\bf Alexander S. Kuleshov${}^{*}$, Anton D. Skripkin${}^{*}$}
{\vskip 0.3cm}
{${}^{*}$ Department of Mechanics and Mathematics, M.~V.~Lomonosov Moscow State University,\\
        Moscow, 119234. \textit{E-mail: kuleshov@mech.math.msu.su, alexander.kuleshov@math.msu.ru}}
\end{center}
{\vskip 1cm}

\begin{abstract}
In 1890 W.~Hess~\cite{Hess} found the new special case of integrability of the Euler --- Poisson equations of motion of a heavy rigid body with a fixed point. In 1963 L.~N.~Sretensky in his paper~\cite{Sretensky} proved that the special case of integrability, similar to the Hess case, also exists in the problem of the motion of a heavy gyrostat --- a heavy rigid body with a fixed point, which contains a rotating homogeneous rotor. Further numerous generalizations of the classical Hess case were proposed~\cite{Golubev,BorisovMamaev1,BorisovMamaev2,GGK,BurovKarapetyan,Lunev,Kozlov,Samsonov,Kholostova,BBM,GorrMaznev,Kosov}, which take place during the motion of a heavy rigid body and a gyrostat with a fixed point in various force fields. The most general conditions of existence of the special case of integrability of the Euler --- Poisson Equations, similar to the Hess case, in the problem of motion of a rigid body with a fixed point and a gyrostat in various force fields, were presented in the paper by A.~A.~Kosov~\cite{Kosov}.

The first studies that provided a qualitative description of the motion of a heavy rigid body in the integrable Hess case were published almost immediately after this case was found. In 1892 P.~A.~Nekrasov proved~\cite{Nekrasov1,Nekrasov2}, that the solution of the problem of motion of a heavy rigid body with a fixed point in the Hess case is reduced to the integration the second order linear homogeneous differential equation with variable coefficients. A similar result regarding the problem of the motion of a heavy gyrostat in the Hess -- Sretensky case was presented by Sretensky~\cite{Sretensky}.

In this paper we present the derivation of the corresponding second order linear differential equation and reduce the coefficients of this equation to the form of rational functions. Then, using the Kovacic algorithm~\cite{Kovacic}, we study the problem of the existence of liouvillian solutions of the corresponding second--order linear differential equation. We obtain the conditions for the parameters of the problem, under which the liouvillian solutions of the corresponding linear differential equation exist. Under these conditions equations of motion of a heavy gyrostat with a fixed point in the Hess -- Sretensky case can be integrated in quadratures.
\end{abstract}


\subsection{Problem formulation. Equations of Motion of a Gyrostat and Its First Integrals.}

Let us consider the problem of motion of a mechanical system $S$, consisting of two joint rigid bodies in a uniform gravitational field. The first body $S_0$ (the carrier) has a point $O$ fixed in the inertial space. The second rigid body, the rotor $S_1$, is an axially symmetric body with the axis of symmetry fixed in the carrier $S_0$. We denote by ${\vec e}$ the unit vector along the axis of symmetry of the rotor $S_1$. The center of mass $O_1$ of the rotor is located on the axis of symmetry. Since the rotor is a rotationally symmetric rigid body, its rotation does not change the position of the center of mass of the system $S$ and the mass distribution of the system. Let us introduce the moving coordinate frame $Ox_1x_2x_3$, which is rigidly connected with the carrier $S_0$. Its axes are directed along the principal axes of inertia of the system at $O$. We denote the unit vectors of $Ox_1x_2x_3$ coordinate system by ${\vec e}_1$, ${\vec e}_2$, ${\vec e}_3$. Let $\mathbb{I}=\diag\left(A_1,\, A_2,\, A_3\right)$ and ${\vec r}_0=x_1{\vec e}_1+x_2{\vec e}_2+x_3{\vec e}_3$ be the inertia matrix and the position vector of the center of mass of the system with respect to $Ox_1x_2x_3$ coordinate frame. Let also $\mathbb{J}$ be the inertia matrix of the rotor with respect to the coordinate frame $O_1x_1y_1z_1$ fixed in it with $O_1z_1$ axis directed along its axis of symmetry. Let us assume that the rotor is set and kept in motion about its axis with constant angular velocity $\Omega$, using some device. Let ${\vec\gamma}=\gamma_1{\vec e}_1+\gamma_2{\vec e}_2+\gamma_3{\vec e}_3$ and ${\vec\omega}=\omega_1{\vec e}_1+\omega_2{\vec e}_2+\omega_3{\vec e}_3$ be the unit vector of upward vertical and the angular velocity vector of the carrier $S_0$ respectively. The angular momentum of the system can be written as follows (see~\cite{Yehia,Wittenburg}):
\begin{equation}\label{1}
{\vec K}=\mathbb{I}{\vec\omega}+\Omega\mathbb{J}{\vec e}.
\end{equation}

The last term in the right-hand side of the last equation ${\vec s}=\Omega\mathbb{J}{\vec e}$ is called the gyrostatic momentum. It is the angular momentum of the rotor $S_1$ relative to the carrier $S_0$. This angular momentum is directed along the axis of symmetry of the rotor. Therefore, we can rewrite~\eqref{1} in the form:
\begin{equation*}
{\vec K}=\mathbb{I}{\vec\omega}+{\vec s}.
\end{equation*}

Now let us derive the equations of motion of the system. The mutual forces between the carrier and the rotor are internal in the system and do not appear in these equations. We have
\begin{equation*}
\dot{{\vec K}}+\left[{\vec\omega}\times{\vec K}\right]=Mg\left[{\vec\gamma}\times{\vec r}_0\right].
\end{equation*}

Since ${\vec s}$ is kept constant in the carrier, $\dot{\vec s}=0$, then the last equation takes the form:
\begin{equation}\label{2}
\mathbb{I}\dot{{\vec\omega}}+\left[{\vec\omega}\times\left(\mathbb{I}{\vec\omega}+{\vec s}\right)\right]=Mg\left[{\vec\gamma}\times{\vec r}_0\right].
\end{equation}

Together with the Poisson's equation
\begin{equation}\label{3}
\dot{{\vec\gamma}}=\left[{\vec\gamma}\times{\vec\omega}\right],
\end{equation}
we obtain a closed system of six first order differential equations:
\begin{equation}\label{4}
\begin{array}{l}
A_1{\dot\omega}_1+\left(A_3-A_2\right)\omega_2\omega_3+s_3\omega_2-s_2\omega_3=Mg\left(x_3\gamma_2-x_2\gamma_3\right),\\ \\
A_2{\dot\omega}_2+\left(A_1-A_3\right)\omega_1\omega_3+s_1\omega_3-s_3\omega_1=Mg\left(x_1\gamma_3-x_3\gamma_1\right),\\ \\
A_3{\dot\omega}_3+\left(A_2-A_1\right)\omega_1\omega_2+s_2\omega_1-s_1\omega_2=Mg\left(x_2\gamma_1-x_1\gamma_2\right),
\end{array}
\end{equation}
\begin{equation}\label{5}
\dot{\gamma}_1=\omega_3\gamma_2-\omega_2\gamma_3,\quad \dot{\gamma}_2=\omega_1\gamma_3-\omega_3\gamma_1,\quad
\dot{\gamma}_3=\omega_2\gamma_1-\omega_1\gamma_2.
\end{equation}

Here $M$ is the mass of the system, $g$ is the acceleration due to gravity. Equations~\eqref{2},~\eqref{3} (or, in scalar form,~\eqref{4},~\eqref{5}) are called the Euler -- Poisson equations of motion of a heavy gyrostat with a fixed point. It is well known~\cite{Yehia,Gavrilov} that to solve the Euler -- Poisson equations we need to find four independent autonomous first integrals of the system~\eqref{4},~\eqref{5}. For any values of parameters $A_1$, $A_2$, $A_3$, $x_1$, $x_2$, $x_3$, $s_1$, $s_2$, $s_3$ of the system and for any initial conditions we have three first integrals of the Euler -- Poisson equations.
\begin{enumerate}
\item The energy integral
\begin{equation*}
\frac{1}{2}\left(A_1\omega_1^2+A_2\omega_2^2+A_3\omega_3^2\right)+
Mg\left(x_1\gamma_1+x_2\gamma_2+x_3\gamma_3\right)=E=\const.
\end{equation*}

\item The area integral
\begin{equation*}
\left(A_1\omega_1+s_1\right)\gamma_1+\left(A_2\omega_2+s_2\right)\gamma_2+\left(A_3\omega_3+s_3\right)\gamma_3=k=\const.
\end{equation*}

\item The geometrical integral
\begin{equation*}
\gamma_1^2+\gamma_2^2+\gamma_3^2=1.
\end{equation*}
\end{enumerate}

Thus, for the integrability of the Euler -- Poisson equations, we need to find only one additional autonomous first integral. In general case the additional integral does not exist. However, under appropriate conditions on the parameters $A_1$, $A_2$, $A_3$, $x_1$, $x_2$, $x_3$, $s_1$, $s_2$, $s_3$, this integral can exist only in the following three cases~\cite{Gavrilov}.
\begin{enumerate}
\item The Euler -- Volterra case (Volterra~\cite{Volterra}, $x_1=x_2=x_3=0$). The additional integral has the form:
\begin{equation*}
L^2=\left(A_1\omega_1+s_1\right)^2+\left(A_2\omega_2+s_2\right)^2+\left(A_3\omega_3+s_3\right)^2=l^2=\const.
\end{equation*}

\item The Lagrange case $A_1=A_2$, $x_1=x_2=0$, $s_1=s_2=0$. In this case the additional integral has the form:
\begin{equation*}
\omega_3=\omega_3^0=\const.
\end{equation*}

\item The Yehia -- Komarov case (Yehia~\cite{Yehia1986}, Komarov~\cite{Komarov}, $A_1=A_2=2A_3$, $x_2=x_3=0$, $s_1=s_2=0$). The additional first integral can be written as follows:
\begin{equation*}
\begin{array}{l}
\left(\omega_1^2-\omega_2^2-\displaystyle\frac{Mgx_1}{A_3}\gamma_1\right)^2+\left(2\omega_1\omega_2-\displaystyle\frac{Mgx_1}{A_3}
\gamma_2\right)^2+\\ \\
+\displaystyle\frac{2s_3}{A_3}\left(\omega_3-\displaystyle\frac{s_3}{A_3}\right)\left(\omega_1^2+\omega_2^2\right)-
\displaystyle\frac{4Mgx_1s_3}{A_3^2}\gamma_3\omega_1=j=\const.
\end{array}
\end{equation*}
\end{enumerate}

There are no other general cases of integrability of Euler -- Poisson equations of a heavy gyrostat~\cite{Gavrilov}. However, there are several cases, when for the special initial conditions we can find the additional special integral. One of these cases is the Hess -- Sretensky case which has been firstly found by Sretensky~\cite{Sretensky}. Let us consider this case in more details. 

\subsection{The Hess -- Sretensky case.}

We suppose that the parameters of the body $A_1$, $A_2$, $A_3$, $x_1$, $x_2$, $x_3$, $s_1$, $s_2$, $s_3$ satisfy the conditions
\begin{equation}\label{6}
s_3=0,\quad x_3=0,\quad A_2\left(A_3-A_1\right)x_2^2=A_1\left(A_2-A_3\right)x_1^2,\quad A_2\geq A_3\geq A_1.
\end{equation}

Let us prove that under conditions~\eqref{6} the Euler -- Poisson equations~\eqref{4},~\eqref{5} have an additional special integral (the Hess -- Sretensky integral) of the form:
\begin{equation}\label{7}
A_1\omega_1 x_1+A_2\omega_2 x_2+\frac{A_1x_1\left(s_2x_1-s_1x_2\right)}{\left(A_3-A_1\right)x_2}=0.
\end{equation}

To prove the existence of the integral~\eqref{7} we write two first equations of the system~\eqref{4} taking into account conditions~\eqref{6}:
\begin{equation}\label{8}
\begin{array}{l}
A_1{\dot\omega}_1+\left(A_3-A_2\right)\omega_2\omega_3-s_2\omega_3=-Mgx_2\gamma_3, \\ \\
A_2{\dot\omega}_2+\left(A_1-A_3\right)\omega_1\omega_3+s_1\omega_3=Mgx_1\gamma_3.
\end{array}
\end{equation}

Multiplying the first of equations~\eqref{8} by $x_1$ and the second by $x_2$ and taking their sum we obtain:
\begin{equation}\label{9}
\frac{d}{dt}\left(A_1\omega_1 x_1+A_2\omega_2 x_2\right)=\left(A_3-A_1\right)\omega_1\omega_3 x_2+\left(A_2-A_3\right)\omega_2\omega_3 x_1+s_2x_1\omega_3-s_1x_2\omega_3.
\end{equation}

Using conditions~\eqref{6} the right hand side of equation~\eqref{9} can be transformed to the form:
\begin{equation}\label{10}
\begin{array}{l}
\left(A_3-A_1\right)\omega_1\omega_3 x_2+\left(A_2-A_3\right)\omega_2\omega_3
x_1+s_2x_1\omega_3-s_1x_2\omega_3=\\ \\
=\displaystyle\frac{\omega_3\left(A_3-A_1\right)x_2}{A_1x_1}\left(A_1\omega_1 x_1+A_2\omega_2 x_2
+\frac{A_1x_1\left(s_2x_1-s_1x_2\right)}{\left(A_3-A_1\right)x_2}\right).
\end{array}
\end{equation}

Taking into account equation~\eqref{10} we can conclude from the equation~\eqref{9}, that if at initial instant the equation~\eqref{7} is valid then it holds during the whole time of motion of the gyrostat. Therefore under conditions~\eqref{6} the Euler -- Poisson equations~\eqref{4},~\eqref{5} possess the special integral~\eqref{7}.

For the first time this case of special integrability of Euler -- Poisson equations of motion of a heavy gyrostat was found by L.N.~Sretensky~\cite{Sretensky}. In the same paper~\cite{Sretensky} Sretensky proved that the the description of motion of a gyrostat in the Hess -- Sretensky case is reduced to the integration of a certain second order linear homogeneous differential equation. Therefore it is possible to set up the problem of existence of liouvillian solutions of the corresponding linear differential equation. For this purpose it is possible to apply the Kovacic algorithm~\cite{Kovacic}, which allows to find liouvillian solutions of a second order linear differential equation in explicit form. If a linear differential equation has no liouvillian solutions, the Kovacic algorithm also allows one to ascertain this fact. The necessary condition for the application of the Kovacic algorithm to a second order linear differential equation is that the coefficients of this equation should be rational functions of argument.

Below we derive the second order linear differential equation in the Hess -- Sretensky case and reduce its coefficients to the form of rational functions. Further we study the problem of existence of liouvillian solutions for the obtained second order linear differential equation using the Kovacic algorithm.

\subsection{Equations of motion of a gyrostat in the special coordinate system}

Let us write equations of motion~\eqref{4},~\eqref{5} of a gyrostat in the Hess -- Sretensky case in the special coordinate system $O\xi_1\xi_2\xi_3$. We shall denote the unit vectors of this coordinate system by ${\vec e}_I$, ${\vec e}_{II}$, ${\vec e}_{III}$. For the first time this coordinate system has been introduced by P.~V.~Kharlamov~\cite{Kharlamov1,Kharlamov2} for the description of motion of a heavy rigid body with a fixed point in the Hess case. The transformation from the principal axes of inertia $Ox_1x_2x_3$ of a gyrostat to the special coordinate system $O\xi_1\xi_2\xi_3$ is defined by the formulas:
\begin{equation*}
{\vec e}_I={\vec e}_1\cos\alpha+{\vec e}_2\sin\alpha, \qquad {\vec e}_{II}=-{\vec e}_1\sin\alpha+{\vec e}_2\cos\alpha,\qquad
{\vec e}_{III}={\vec e}_3,
\end{equation*}
where $\cos\alpha$ and $\sin\alpha$ equal
\begin{equation*}
\cos\alpha=\frac{x_1}{\sqrt{x_1^2+x_2^2}}, \qquad \sin\alpha=\frac{x_2}{\sqrt{x_1^2+x_2^2}}.
\end{equation*}

In the $O\xi_1\xi_2\xi_3$ coordinate system the angular momentum ${\vec K}$ of a gyrostat can be written as follows:
\begin{equation*}
{\vec K}=\left(L_1+k_1\right){\vec e}_I+\left(L_2+k_2\right){\vec e}_{II}+L_3{\vec e}_{III}.
\end{equation*}

Here the components $L_1$, $L_2$, $L_3$ of the angular momentum of a carrier $S_0$ and the components $k_1$ and $k_2$ of the angular momentum of a rotor $S_1$ are connected with the components $A_1\omega_1$, $A_2\omega_2$, $A_3\omega_3$, $s_1$, $s_2$ by formulas
\begin{equation*}
\begin{array}{l}
L_1=\displaystyle\frac{A_1\omega_1x_1+A_2\omega_2x_2}{\sqrt{x_1^2+x_2^2}},\quad L_2=\displaystyle\frac{A_2\omega_2x_1-A_1\omega_1x_2}{\sqrt{x_1^2+x_2^2}},\quad
L_3=A_3\omega_3,\\ \\
k_1=\displaystyle\frac{s_1x_1+s_2x_2}{\sqrt{x_1^2+x_2^2}},\quad k_2=\displaystyle\frac{s_2x_1-s_1x_2}{\sqrt{x_1^2+x_2^2}}.
\end{array}
\end{equation*}

In the $O\xi_1\xi_2\xi_3$ coordinate system the unit vector $\vec{\gamma}$ has the components $\nu_1$, $\nu_2$, $\nu_3$, which are connected with $\gamma_1$, $\gamma_2$, $\gamma_3$ by formulas
\begin{equation*}
\nu_1=\displaystyle\frac{\gamma_1x_1+\gamma_2x_2}{\sqrt{x_1^2+x_2^2}},\quad
\nu_2=\displaystyle\frac{\gamma_2x_1-\gamma_1x_2}{\sqrt{x_1^2+x_2^2}},\quad
\nu_3=\gamma_3.
\end{equation*}

Using the variables $L_1$, $L_2$, $L_3$, $\nu_1$, $\nu_2$, $\nu_3$ we can rewrite the Euler --- Poisson equations~\eqref{4},~\eqref{5} in the form
\begin{equation}\label{11}
\begin{array}{l}
\dot{L}_1=-bL_3\left(L_1-\displaystyle\frac{ck_2}{b}\right),\quad \dot{L}_2=\left(a-c\right)L_1L_3+bL_2L_3-ck_1L_3+\nu_3\Gamma,
\\ \\
\dot{L}_3=-\left(a-c\right)L_1L_2+bL_1^2-bL_2^2+\left(k_1b-k_2a\right)L_1+\left(k_1c-k_2b\right)L_2-\nu_2\Gamma,\\ \\
\dot{\nu}_1=cL_3\nu_2-\left(cL_2+bL_1\right)\nu_3,\quad \dot{\nu}_2=\left(aL_1+bL_2\right)\nu_3-cL_3\nu_1,\\ \\
\dot{\nu}_3=-\left(aL_1+bL_2\right)\nu_2+\left(cL_2+bL_1\right)\nu_1.
\end{array}
\end{equation}

Here we introduce the following parameters
\begin{equation*}
a=\frac{A_2x_1^2+A_1x_2^2}{A_1A_2\left(x_1^2+x_2^2\right)},\quad
b=\frac{\left(A_1-A_2\right)x_1x_2}{A_1A_2\left(x_1^2+x_2^2\right)},\quad c=\frac{1}{A_3},\quad \Gamma=Mg\sqrt{x_1^2+x_2^2}.
\end{equation*}

To find the additional first integral, existing in the Hess -- Sretensky case, we consider the first equation of the system~\eqref{11}
\begin{equation*}
\dot{L}_1=-bL_3\left(L_1-\displaystyle\frac{ck_2}{b}\right)
\end{equation*}

In this equation, the right hand side is equal to the expression 
\begin{equation*}
L_1-\displaystyle\frac{ck_2}{b},
\end{equation*}
multiplied by the coefficient $-bL_3$ bounded in absolute value. This means that if at the initial instant of time 
we have
\begin{equation*}
L_1=\displaystyle\frac{ck_2}{b},
\end{equation*}
then we have
\begin{equation}\label{12}
L_1\equiv\displaystyle\frac{ck_2}{b}
\end{equation}
during the whole time of motion of a gyrostat. 

The invariant manifold~\eqref{12} (or, in other notations~\eqref{7}) together with the conditions~\eqref{6} defines the Hess -- Sretensky integrable case of motion of a heavy gyrostat with a fixed point. Under conditions~\eqref{6},~\eqref{12} equations~\eqref{11} are noticeably simplified and take the form
\begin{equation}\label{13}
\begin{array}{l}
\dot{\tilde{L}}_2=b\tilde{L}_2L_3+\left(\displaystyle\frac{\left(ac-b^2\right)k_2}{b}-c\left(\displaystyle\frac{ck_2}{b}+k_1\right)
\right)L_3+\nu_3\Gamma,\\ \\ 
\dot{L}_3=-b\tilde{L}_2^2-\left(\displaystyle\frac{\left(ac-b^2\right)k_2}{b}-c\left(\displaystyle\frac{ck_2}{b}+k_1\right)
\right)\tilde{L}_2-\nu_2\Gamma,\\ \\
\dot{\nu}_1=cL_3\nu_2-c\tilde{L}_2\nu_3,\quad \dot{\nu}_2=-cL_3\nu_1+b\tilde{L}_2\nu_3+\displaystyle\frac{\left(ac-b^2\right)k_2}{b}\nu_3,\\ \\
\dot{\nu}_3=c\tilde{L}_2\nu_1-b\tilde{L}_2\nu_2-\displaystyle\frac{\left(ac-b^2\right)k_2}{b}\nu_2.
\end{array}
\end{equation}

Here we denote by $\tilde{L}_2$ the following expression 
\begin{equation*}
\tilde{L}_2=L_2+k_2.
\end{equation*}

Equations~\eqref{13} admit three first integrals:
\begin{equation}\label{14}
\displaystyle\frac{c}{2}\left(\tilde{L}_2^2+L_3^2\right)+\Gamma\nu_1=E,\quad
\tilde{L}_2\nu_2+L_3\nu_3+\left(\displaystyle\frac{ck_2}{b}+k_1\right)\nu_1=k,\quad
\nu_1^2+\nu_2^2+\nu_3^2=1.
\end{equation}

\subsection{Dimensionless equations. Transformation to the second-order linear differential equation}

Now let us write the equations~\eqref{13} and the first integrals~\eqref{14} in dimensionless form. For this purpose we firstly introduce the following notations
\begin{equation*}
F=\frac{\left(ac-b^2\right)k_2}{b},\quad G=\frac{ck_2}{b}+k_1.
\end{equation*}

We can rewrite now equations~\eqref{13} and the first integrals~\eqref{14} in the form:
\begin{equation}\label{15}
\begin{array}{l}
\dot{\tilde{L}}_2=b\tilde{L}_2L_3+\left(F-Gc\right)L_3+\nu_3\Gamma,\quad \dot{L}_3=-b\tilde{L}_2^2-\left(F-Gc\right)\tilde{L}_2-\nu_2\Gamma,\\ \\
\dot{\nu}_1=cL_3\nu_2-c\tilde{L}_2\nu_3,\; \dot{\nu}_2=-cL_3\nu_1+b\tilde{L}_2\nu_3+F\nu_3,\;
\dot{\nu}_3=c\tilde{L}_2\nu_1-b\tilde{L}_2\nu_2-F\nu_2,
\end{array}
\end{equation}
\begin{equation}\label{16}
\begin{array}{l}
\displaystyle\frac{c}{2}\left(\tilde{L}_2^2+L_3^2\right)+\Gamma\nu_1=E;\quad G\nu_1+\tilde{L}_2\nu_2+L_3\nu_3=k;\quad
\nu_1^2+\nu_2^2+\nu_3^2=1.
\end{array}
\end{equation}

Let us introduce the dimensionless components of angular momentum
\begin{equation*}
\tilde{L}_2=\sqrt{\frac{\Gamma}{c}}y,\qquad L_3=\sqrt{\frac{\Gamma}{c}}z,
\end{equation*}
the dimensionless time $\tau$:
\begin{equation*}
t=\frac{\tau}{\sqrt{\Gamma c}},
\end{equation*}
the dimensionless constants of the first integrals
\begin{equation*}
h=\frac{E}{\Gamma},\qquad p_1=k\sqrt{\frac{c}{\Gamma}}
\end{equation*}
and the dimensionless parameters
\begin{equation*}
d_1=\frac{b}{c},\quad A=\displaystyle\frac{F}{\sqrt{\Gamma c}},\quad B=G\sqrt{\displaystyle\frac{c}{\Gamma}}.
\end{equation*}

Now we can rewrite equations~\eqref{15} and the first integrals~\eqref{16} in dimensionless form:
\begin{equation}\label{17}
\begin{array}{l}
\displaystyle\frac{dy}{d\tau}=d_1yz+\left(A-B\right)z+\nu_3,\quad \displaystyle\frac{dz}{d\tau}=-d_1y^2-\left(A-B\right)y-\nu_2,\\ \\
\displaystyle\frac{d{\nu}_1}{d\tau}=z\nu_2-y\nu_3,\quad \displaystyle\frac{d{\nu}_2}{d\tau}=d_1y\nu_3-z\nu_1+A\nu_3,\quad
\displaystyle\frac{d{\nu}_3}{d\tau}=-d_1y\nu_2+y\nu_1-A\nu_2,
\end{array}
\end{equation}
\begin{equation}\label{18}
\displaystyle\frac{y^2+z^2}{2}+\nu_1=h,\quad y\nu_2+z\nu_3+B\nu_1=p_1,\quad \nu_1^2+\nu_2^2+\nu_3^2=1.
\end{equation}

From the system~\eqref{17} using the first integrals~\eqref{18} we shall obtain the second order linear differential equation. Before we obtain this equation, let us determine the range of parameters $A$, $B$, $d_1$, $p_1$, $h$. It is easy to see that the parameters $A$, $B$ and $p_1$ ranges in the infinite interval $\left(-\infty,\, +\infty\right)$. Since the expression
\begin{equation*}
\frac{y^2+z^2}{2}
\end{equation*}
is nonnegative, then we have for the parameter $h$ the following inequality
\begin{equation*}
h-\nu_1\geq 0,\quad\mbox{or}\quad h\geq\nu_1.
\end{equation*}

The minimal value of the first component $\nu_1$ of the vector $\vec\nu$ equals $-1$. Therefore the parameter $h$ ranges in the interval
\begin{equation*}
h\in\left[-1,\;+\infty\right).
\end{equation*}

It can be shown (see~\cite{BardinKuleshov1,BardinKuleshov2}) that the parameter $d_1$ ranges in the interval:
\begin{equation*}
d_1\in\left(-1,\, 0\right].
\end{equation*}

We obtain now the second order linear differential equation from the system~\eqref{17} using the first integrals~\eqref{18}. Multiplying the first equation of the system~\eqref{17} by $y$ and the second by $z$ and adding them, we get:
\begin{equation}\label{19}
\frac{d}{d\tau}\left(\frac{y^2+z^2}{2}\right)=y\nu_3-z\nu_2.
\end{equation}

From the first integrals~\eqref{18} we have the following equations:
\begin{equation}\label{20}
\begin{array}{l}
\nu_1=h-\displaystyle\frac{y^2+z^2}{2},\quad \nu_2^2+\nu_3^2=1-\nu_1^2=1-\left(\displaystyle\frac{y^2+z^2}{2}-h\right)^2,\\ \\
y\nu_2+z\nu_3=p_1-B\nu_1=p_1-Bh+\displaystyle\frac{B}{2}\left(y^2+z^2\right).
\end{array}
\end{equation}

Therefore, from the trivial identity
\begin{equation*}
\left(y^2+z^2\right)\left(\nu_2^2+\nu_3^2\right)=\left(y\nu_2+z\nu_3\right)^2+\left(y\nu_3-z\nu_2\right)^2,
\end{equation*}
we obtain:
\begin{equation*}
\left(y\nu_3-z\nu_2\right)^2=\left(y^2+z^2\right)\left(1-\left(\frac{y^2+z^2}{2}-h\right)^2\right)-\left(p_1-Bh+
\displaystyle\frac{B}{2}\left(y^2+z^2\right)\right)^2.
\end{equation*}

We will take that
\begin{equation}\label{21}
y\nu_3-z\nu_2=-\sqrt{\left(y^2+z^2\right)\left(1-\left(\frac{y^2+z^2}{2}-h\right)^2\right)-\left(p_1-Bh+
\displaystyle\frac{B}{2}\left(y^2+z^2\right)\right)^2}
\end{equation}
(we can choose any of the sign before the square root in~\eqref{21}). Taking into account~\eqref{21} we can write equation~\eqref{19} as follows:
\begin{equation*}
\displaystyle\frac{d}{d\tau}\left(\displaystyle\frac{y^2+z^2}{2}\right)=
-\sqrt{\left(y^2+z^2\right)\left(1-\left(\displaystyle\frac{y^2+z^2}{2}-h\right)^2\right)-\left(p_1-Bh+
\displaystyle\frac{B}{2}\left(y^2+z^2\right)\right)^2}.
\end{equation*}

Now we multiply the second equation of the system~\eqref{17} by $y$ and the first by $-z$ and add them. As a result, we obtain:
\begin{equation*}
y\displaystyle\frac{dz}{d\tau}-z\displaystyle\frac{dy}{d\tau}=-d_1y\left(y^2+z^2\right)-\left(A-B\right)\left(y^2+z^2\right)-
p_1+Bh-\displaystyle\frac{B}{2}\left(y^2+z^2\right).
\end{equation*}

We introduce now the new variables $x$ and $\varphi$ ("polar coordinates") such that
\begin{equation*}
y=x\cos\varphi,\quad z=x\sin\varphi.
\end{equation*}

Then for the variables $x$ and $\varphi$ we have the following system of two differential equations:
\begin{equation}\label{22}
\begin{array}{l}
x\displaystyle\frac{dx}{d\tau}=-\sqrt{x^2\left(1-\left(\displaystyle\frac{x^2}{2}-h\right)^2\right)-
\left(p_1+B\left(\displaystyle\frac{x^2}{2}-h\right)\right)^2}, \\ \\
x^2\displaystyle\frac{d\varphi}{d\tau}=-d_1x^3\cos\varphi-\left(A-B\right)x^2-p_1-B\left(\displaystyle\frac{x^2}{2}-h\right).
\end{array}
\end{equation}

From the system~\eqref{22} we obtain the single first order differential equation for the function $\varphi=\varphi\left(x\right)$:
\begin{equation}\label{23}
\frac{d\varphi}{dx}=\frac{d_1x^3\cos\varphi+\left(A-\displaystyle\frac{B}{2}\right)x^2-\left(p_1-Bh\right)}{x\sqrt{x^2\left(1-\left(
\displaystyle\frac{x^2}{2}-h\right)^2\right)-\left(p_1-Bh+\displaystyle\frac{B}{2}x^2\right)^2}}.
\end{equation}

Note that when we obtain the equation~\eqref{23} from the system~\eqref{22} we exclude the case $x={\rm const}$ that is, $\nu_1={\rm const}$ from consideration. Meanwhile for a heavy rigid body and a gyrostat with a fixed point in the Hess -- Sretensky case there are steady motions for which $\nu_1=\nu_1^0={\rm const}$ (see, for example~\cite{Novikov}).

Using the change of variables
\begin{equation*}
w=\tan\frac{\varphi}{2}
\end{equation*}
we reduce the equation~\eqref{23} to the Riccati equation:
\begin{equation}\label{24}
\frac{dw}{dx}=f_2w^2+f_0,
\end{equation}
\begin{equation*}
\begin{array}{l}
f_2=-\displaystyle\frac{2d_1x^3+\left(B-2A\right)x^2+2\left(Bh-p_1\right)}{2x
\sqrt{4\left(1-h^2+B^2h-p_1B\right)x^2+\left(4h-B^2\right)x^4-x^6-4\left(p_1-Bh\right)^2}},\\ \\
f_0=\displaystyle\frac{2d_1x^3+\left(2A-B\right)x^2+2\left(p_1-Bh\right)}
{2x\sqrt{4\left(1-h^2+B^2h-p_1B\right)x^2+\left(4h-B^2\right)x^4-x^6-4\left(p_1-Bh\right)^2}}.
\end{array}
\end{equation*}

It is well known from the general theory of ordinary differential equations (see, for example~\cite{ZaitsevPolyanin}), that if the Riccati equation has the form:
\begin{equation*}
\frac{dw}{dx}=f_2w^2+f_0,
\end{equation*}
then the substitution
\begin{equation*}
u\left(x\right)=\exp\left(-\int f_2 w dx\right)
\end{equation*}
reduces it to the second order linear differential equation
\begin{equation}\label{25}
f_2\frac{d^2u}{dx^2}-\frac{df_2}{dx}\frac{du}{dx}+f_0f_2^2u=0,
\end{equation}
or, if we divide this equation by $f_2$:
\begin{equation}\label{26}
\frac{d^2u}{dx^2}-\frac{1}{f_2}\frac{df_2}{dx}\frac{du}{dx}+f_0f_2u=0.
\end{equation}

Note that the transition from the equation~\eqref{25} to the equation~\eqref{26} is possible only when $f_2\ne 0$. Taking into account the fact that $x\ne {\rm const}$, the condition $f_2=0$ is equivalent to the simultaneous fulfillment of the conditions
\begin{equation*}
d_1=0,\quad B=2A,\quad p_1=Bh.
\end{equation*}

Further we shall assume that $f_2\ne 0$. Moreover we shall assume that $d_1\ne 0$. Thus, the following theorem is valid.
\begin{thm}\label{Theorem1}
The problem of motion of a heavy gyrostat with a fixed point in the Hess -- Sretensky case is reduced to solving the second order linear homogeneous differential equation with rational coefficients. $\Box$
\end{thm}

In the explicit form this linear differential equation can be written as follows:
\begin{equation}\label{27}
\frac{d^2u}{dx^2}+a\left(x\right)\frac{du}{dx}+b\left(x\right)u=0,
\end{equation}
\begin{equation*}
a\left(x\right)=\frac{P_9\left(x\right)}{xP_6\left(x\right)P_3\left(x\right)},\quad b\left(x\right)=\frac{P_3\left(x\right)Q_3\left(x\right)}{4x^2P_6\left(x\right)},
\end{equation*}

\begin{equation*}
P_3\left(x\right)=2d_1x^3+\left(B-2A\right)x^2+2\left(Bh-p_1\right),\quad Q_3\left(x\right)=4d_1x^3-P_3\left(x\right),
\end{equation*}

\begin{equation*}
P_6\left(x\right)=x^6+\left(B^2-4h\right)x^4+4\left(h^2-1-B\left(Bh-p_1\right)\right)x^2+4\left(Bh-p_1\right)^2,
\end{equation*}

\begin{equation*}
\begin{array}{l}
P_9\left(x\right)=2d_1x^9+2\left(B-2A\right)x^8+\left(B^3+8Ah-2AB^2+4Bh-8p_1\right)x^6+\\ \\
+8d_1\left(1-h^2-p_1B+hB^2\right)x^5\!+\!6\left(B^2-4h\right)\left(Bh\!-\!p_1\right)x^4\!-\!16d_1\left(Bh\!-\!p_1\right)^2x^3+\\ \\
+\left(4\left(Bh-p_1\right)^2\left(2A-5B\right)+16\left(Bh-p_1\right)\left(h^2-1\right)\right)x^2+8\left(Bh-p_1\right)^3.
\end{array}
\end{equation*}

Since the coefficients of the obtained second order linear differential equation~\eqref{27} are rational functions of $x$, therefore, we can use the Kovacic algorithm~\cite{Kovacic} to find liouvillian solutions of this differential equation. Below we give a brief description of the Kovacic algorithm. 

\subsection{Description of the Kovacic algorithm}

Let us consider the differential field $\mathbb{C}\left(x\right)$ of rational functions of one (in general case complex) variable $x$. We accept the standard notations $\mathbb{Z}$ and $\mathbb{Q}$ for the sets of integer and rational numbers respectively. Our goal is to find a solution of the differential equation
\begin{equation}\label{28}
\frac{d^2z}{dx^2}+a\left(x\right)\frac{dz}{dx}+b\left(x\right)z=0,
\end{equation}
where $a\left(x\right), b\left(x\right)\in \mathbb{C}\left(x\right)$. In the paper~\cite{Kovacic} an algorithm has been proposed that allows one to find explicitly the so-called liouvillian solutions of differential equation~\eqref{28}, i.e. solutions, that can be expressed in terms of liouvillian functions. The main advantage of the Kovacic algorithm is precisely that it allows one not only to establish the existence or nonexistence of a solution of differential equation~\eqref{28} expressed in terms of liouvillian functions, but also to present this solution in an explicit form when it exists. In turn, liouvillian functions are elements of a liouvillian field, which is defined in the following way.

\begin{Def}
Let $F$ be a differential field of functions of one (in general case complex) variable $x$ that contains $\mathbb{C}\left(x\right)$; namely $F$ is a field of characteristic zero with a differentiation operator $\left(\right)'$ with the following two properties: $\left(a+b\right)'=a'+b'$ and $\left(ab\right)'=a'b+ab'$ for any $a$ and $b$ in $F$. The field $F$ is liouvillian if there exists a sequence (tower) of differential fields
\begin{equation*}
\mathbb{C}\left(x\right)=F_0\subseteq F_1\subseteq\ldots\subseteq F_n=F,
\end{equation*}
obtained by adjoining one element such that for any $i~=~1,2,\ldots, n$ we have:
\begin{equation*}
F_i=F_{i-1}\left(\alpha\right),\quad\mbox{with}\quad \frac{\alpha'}{\alpha}\in F_{i-1}
\end{equation*}
(i.e. $F_i$ is generated by an exponential of an indefinite integral over $F_{i-1}$); or
\begin{equation*}
F_i=F_{i-1}\left(\alpha\right),\quad\mbox{with}\quad \alpha'\in F_{i-1}
\end{equation*}
(i.e. $F_i$ is generated by an indefinite integral over $F_{i-1}$); or $F_i$ is finite algebraic over $F_{i-1}$ (i.e. $F_i=F_{i-1}\left(\alpha\right)$ and $\alpha$ satisfies a polynomial equation of the form
\begin{equation*}
a_0+a_1\alpha+\cdots+a_n\alpha^n=0,
\end{equation*}
where $a_j\in F_{i-1}$, $j=0,1,2,\ldots, n$ and are not all zero). $\Box$
\end{Def}

Thus, liouvillian functions are built up sequentially from rational functions by using algebraic operations and the operation of indefinite integration and by taking the exponential of a given expression. A solution of equation~\eqref{28} that is expressed in terms of liouvillian functions most closely correspond to the notion of a "close-form solution"$\,$ or a "solution in quadratures".
To reduce differential equation~\eqref{28} to a simpler form, we use the following formula
\begin{equation}\label{29}
y\left(x\right)=z\left(x\right)\exp\left(\frac{1}{2}\int a(x)dx\right).
\end{equation}

Then equation~\eqref{28} takes the form:
\begin{equation}\label{30}
y''=R\left(x\right)y,\quad R\left(x\right)=\frac{1}{2}a'+\frac{1}{4}a^2-b,\quad R\left(x\right)\in\mathbb{C}\left(x\right).
\end{equation}

Hereinafter, it is assumed that that the second order linear differential equation with which the Kovacic algorithm deals is written in the form~\eqref{30}. The following theorem which has been proved by J.~Kovacic~\cite{Kovacic}, determines the structure of a solution of this differential equation.
\begin{thm}\label{Theorem2}
For the differential equation~\eqref{30} only the following four cases are true.
\begin{enumerate}
\item The differential equation~\eqref{30} has a solution of the form
\begin{equation*}
\eta=\exp\left(\int\omega(x)dx\right)\quad\mbox{e}\quad\omega\left(x\right)\in\mathbb{C}\left(x\right)
\end{equation*}
(liouvillian solution of type $1$).

\item The differential equation~\eqref{30} has a solution of the form
\begin{equation*}
\eta=\exp\left(\int\omega(x)dx\right),
\end{equation*}
where $\omega\left(x\right)$ is an algebraic function of degree $2$ over $\mathbb{C}\left(x\right)$ and case $1$ does not hold (liouvillian solution of type $2$).

\item All solutions of differential equation~\eqref{30} are algebraic over $\mathbb{C}\left(x\right)$ and cases $1$ and $2$ do not hold. In this situation a solution of the differential equation~\eqref{30} has the form
\begin{equation*}
\eta=\exp\left(\int\omega(x)dx\right)
\end{equation*}
where $\omega\left(x\right)$ is an algebraic function of degree $4$, $6$ or $12$ over $\mathbb{C}\left(x\right)$ (liouvillian solution of type $3$).

\item Differential equation~\eqref{30} has no liouvillian solutions. $\Box$
\end{enumerate}
\end{thm}

In order for one of the first three cases listed in Theorem~\ref{Theorem2} to take place the function $R\left(x\right)$ in the right hand side of equation~\eqref{30} must satisfy certain conditions. These conditions are necessary but not sufficient. For example, if the conditions corresponding to Case 1 of Theorem~\ref{Theorem2} are violated, then we must turn to the verification of the conditions corresponding to Cases 2 and 3. If these conditions are fulfilled, then we must search for solutions of equation~\eqref{30} exactly in the form, indicated for the corresponding case. However, the existence of such a solution is not guaranteed. In order to explain the sense of the necessary conditions mentioned, we recall some facts from complex analysis.

Recall that any analytic function $f$ of a complex variable $z$ can be expanded in a Laurent series in a neighborhood of any point $a$ as follows:
\begin{equation*}
f\left(z\right)=a_0+a_1\left(z-a\right)+a_2\left(z-a\right)^2+\cdots+\frac{a_{-1}}{z-a}+\frac{a_{-2}}{\left(z-a\right)^2}+\cdots.
\end{equation*}

The part of this series
\begin{equation*}
a_0+a_1\left(z-a\right)+a_2\left(z-a\right)^2+\cdots
\end{equation*}
containing nonnegative powers of $z-a$ is called the analytic part of the Laurent series. Whereas the other part, namely
\begin{equation*}
\frac{a_{-1}}{z-a}+\frac{a_{-2}}{\left(z-a\right)^2}+\cdots
\end{equation*}
is called the principal part of the expansion. By definition, a point $a$ is called a pole of $f\left(z\right)$ of order $n$ if the principal part of the Laurent expansion contains a finite number of terms and the last term has the form
\begin{equation*}
\frac{a_{-n}}{\left(z-a\right)^n}.
\end{equation*}

If $f\left(z\right)$ is a rational function of $z$, then a point $a$ is a pole of $f\left(z\right)$ of order $n$ if it is a root of the denominator of $f\left(z\right)$ of multiplicity $n$.

Let $z=\infty$ be a zero of a function $f\left(z\right)$ of order $n$ (i.e., $n$ is the order of the pole at $z=0$ of $f\left(z\right)$). Then we say that $n$ is the order of $f\left(z\right)$ at $z=\infty$. If $f\left(z\right)$ is a rational function, then its order at $z=\infty$ is the difference between the degrees of the denominator and the numerator.

The following theorem, which has been proved in~\cite{Kovacic}, specifies conditions, that are necessary for one of the first three cases listed in Theorem~\ref{Theorem2} can hold.

\begin{thm}\label{Theorem3}
For the differential equation~\eqref{30} the following conditions are necessary for one of the first cases listed in Theorem~\ref{Theorem2} to hold, i.e. for equation~\eqref{30} to have a liouvillian solution of the type specified in description of the corresponding case.
\begin{enumerate}
\item Each pole of the function $R\left(x\right)$ must have even order or else have order $1$. The order of $R\left(x\right)$ at $x=\infty$ must be even or else be greater than 2.
\item The function $R\left(x\right)$ must have at least one pole that either has odd order greater than $2$ or else has order $2$.
\item The order of a pole of $R\left(x\right)$ cannot exceed $2$ and the order of $R\left(x\right)$ at $x=\infty$ must be at least $2$. If the partial fraction expansion of $R\left(x\right)$ has the form
\begin{equation*}
R\left(x\right)=\sum\limits_i\frac{\alpha_i}{\left(x-c_i\right)^2}+\sum\limits_j\frac{\beta_j}{x-d_j},
\end{equation*}
then for each $i$
\begin{equation*}
\sqrt{1+4\alpha_i}\in\mathbb{Q},\quad \sum\limits_j\beta_j=0
\end{equation*}
and if
\begin{equation*}
\gamma=\sum\limits_i\alpha_i+\sum\limits_j\beta_j d_j,
\end{equation*}
then
\begin{equation*}
\sqrt{1+4\gamma}\in\mathbb{Q}.\quad \Box
\end{equation*}
\end{enumerate}
\end{thm}

To find a liouvillian solution of type 1 of the differential equation~\eqref{30}, the Kovacic algorithm is stated in the following way (see~\cite{Kovacic,BardinKuleshov1} for details). We assume that the necessary conditions for the existence of a solution in case 1 are satisfied and denote the set of finite poles of the function $R\left(x\right)$ by $\Gamma$.
\begin{description}
\item[\underline{Step 1.}] For each $c\in\Gamma\bigcup\left\{\infty\right\}$ we define a rational function $\left[\sqrt{R}\right]_c$ and two complex numbers $\alpha_{c}^{+}$ and $\alpha_{c}^{-}$ as described below.
\begin{itemize}
\item[$\left(c_1\right)$] If $c\in\Gamma$ is a pole of order 1, then
\begin{equation*}
\left[\sqrt{R}\right]_c=0,\quad \alpha_{c}^{+}=\alpha_{c}^{-}=1.
\end{equation*}
\item[$\left(c_2\right)$] If $c\in\Gamma$ is a pole of order 2, then
\begin{equation*}
\left[\sqrt{R}\right]_c=0.
\end{equation*}

Let $b$ be the coefficient of $\displaystyle\frac{1}{\left(x-c\right)^2}$ in the partial fraction expansion of $R\left(x\right)$. Then
\begin{equation*}
\alpha_{c}^{\pm}=\frac{1}{2}\pm\frac{1}{2}\sqrt{1+4b}.
\end{equation*}
\item[$\left(c_3\right)$] If $c\in\Gamma$ is a pole of order $2\nu\geq 4$ (the order must be even due to the necessary conditions, stated in Theorem~\ref{Theorem3}), then $\left[\sqrt{R}\right]_c$ is the sum of terms involving $\displaystyle\frac{1}{\left(x-c\right)^i}$ for $2\leq i\leq\nu$ in the Laurent expansion of $\sqrt{R}$ at $c$. There are two possibilities for $\left[\sqrt{R}\right]_c$ that differ by sign; we can choose one of them. Thus,
\begin{equation*}
\left[\sqrt{R}\right]_c=\frac{a}{\left(x-c\right)^{\nu}}+\cdots+\frac{d}{\left(x-c\right)^2}.
\end{equation*}

Let $b$ be the coefficient of $\displaystyle\frac{1}{\left(x-c\right)^{\nu+1}}$ in $R-\left[\sqrt{R}\right]^2_c$. Then
\begin{equation*}
\alpha_{c}^{\pm}=\frac{1}{2}\left(\pm\frac{b}{a}+\nu\right).
\end{equation*}
\item[$\left(\infty_1\right)$] If the order of $R\left(x\right)$ at $x=\infty$ is greater than 2, then
\begin{equation*}
\left[\sqrt{R}\right]_{\infty}=0,\quad \alpha_{\infty}^{+}=1,\quad \alpha_{\infty}^{-}=0.
\end{equation*}
\item[$\left(\infty_2\right)$] If the order of $R\left(x\right)$ at $x=\infty$ is 2, then
\begin{equation*}
\left[\sqrt{R}\right]_{\infty}=0.
\end{equation*}

Let $b$ be the coefficient of $\displaystyle\frac{1}{x^2}$ in the Laurent series expansion of $R\left(x\right)$ at $x=\infty$. Then
\begin{equation*}
\alpha_{\infty}^{\pm}=\frac{1}{2}\pm\frac{1}{2}\sqrt{1+4b}.
\end{equation*}
\item[$\left(\infty_3\right)$] If the order $R\left(x\right)$ at $x=\infty$ is $-2\nu\leq 0$ (it is even due to the necessary conditions stated in Theorem~\ref{Theorem3}), then the function $\left[\sqrt{R}\right]_{\infty}$ is the sum of terms involving $x^i$, $0\leq i\leq\nu$ of the Laurent expansion of $\sqrt{R}$ at $x=\infty$ (one of the two possibilities can be chosen). Thus,
\begin{equation*}
\left[\sqrt{R}\right]_{\infty}=ax^{\nu}+\cdots+d.
\end{equation*}

Let $b$ be the coefficient of $x^{\nu-1}$ in $R-\left(\left[\sqrt{R}\right]_{\infty}\right)^2$. Then we have:
\begin{equation*}
\alpha_{\infty}^{\pm}=\frac{1}{2}\left(\pm\frac{b}{a}-\nu\right).
\end{equation*}
\end{itemize}

\item[\underline{Step 2.}] For each family $s=\left(s\left(c\right)\right)_{c\in\Gamma\bigcup\left\{\infty\right\}}$, where $s\left(c\right)$ are either $+$ or $-$ let
\begin{equation}\label{31}
d=\alpha_{\infty}^{s\left(\infty\right)}-\sum\limits_{c\in\Gamma}\alpha_c^{s\left(c\right)}.
\end{equation}

If $d$ is a non -- negative integer, then we introduce the function
\begin{equation}\label{32}
\theta=\sum\limits_{c\in\Gamma}\left(s\left(c\right)\left[\sqrt{R}\right]_{c}+\frac{\alpha_{c}^{s\left(c\right)}}{x-c}\right)
+s\left(\infty\right)\left[\sqrt{R}\right]_{\infty}
\end{equation}

If $d$ is not a non -- negative integer, then the family $s$ should be discarded. If all tuples $s$ have been rejected, then Case~1 cannot hold.

\item[\underline{Step 3.}] For each family $s$ from Step~2, we search for a monic polynomial $P$ of degree $d$ (the constant $d$ is defined by the formula~\eqref{31}), satisfying the differential equation
\begin{equation}\label{33}
P''+2\theta P'+\left(\theta'+\theta^2-R\right)P=0.
\end{equation}

If such a polynomial exists, then
\begin{equation*}
\eta=P\exp\left(\int\theta\left(x\right)dx\right)
\end{equation*}
is the solution of the differential equation~\eqref{30}. If for each tuple $s$ found on Step 2, we cannot find such
a polynomial $P$, then Case 1 cannot hold for the differential equation~\eqref{30}.
\end{description}

Now we state the Kovacic algorithm to search for a solution of type 2 of differential equation~\eqref{30}. We denote the set of finite poles of the function $R\left(x\right)$ by $\Gamma$.
\begin{description}
\item[\underline{Step 1.}] For each $c\in\Gamma\bigcup\left\{\infty\right\}$ we define the set $E_c$ as follows.
\begin{itemize}
\item[$\left(c_1\right)$] If $c\in\Gamma$ is a pole of order 1, then
\begin{equation*}
E_c=\left\{4\right\}.
\end{equation*}
\item[$\left(c_2\right)$] If $c\in\Gamma$ is a pole of order 2 and if $b$ is the coefficient of $\frac{1}{\left(x-c\right)^2}$ in the partial fraction expansion of $R\left(x\right)$, then
\begin{equation*}
E_c=\left\{\left(2+k\sqrt{1+4b}\right)\bigcap\mathbb{Z}\right\},\; k=0, \pm2.
\end{equation*}

\item[$\left(c_3\right)$] If $c\in\Gamma$ is a pole of order $\nu>2$, then
\begin{equation*}
E_c=\left\{\nu\right\}.
\end{equation*}
\item[$\left(\infty_1\right)$] If $R\left(x\right)$ has order $>2$ at $x=\infty$, then
\begin{equation*}
E_{\infty}=\left\{0, 2, 4\right\}.
\end{equation*}

\item[$\left(\infty_2\right)$] If $R\left(x\right)$ has order 2 at $x=\infty$ and $b$ is the coefficient of $\frac{1}{x^2}$ in the Laurent series expansion of $R$ at $x=\infty$, then
\begin{equation*}
E_{\infty}=\left\{\left(2+k\sqrt{1+4b}\right)\bigcap\mathbb{Z}\right\},\; k=0, \pm2.
\end{equation*}

\item[$\left(\infty_3\right)$] If $R\left(x\right)$ has order $\nu<2$ at $x=\infty$, then
\begin{equation*}
E_{\infty}=\left\{\nu\right\}.
\end{equation*}
\end{itemize}

\item[\underline{Step 2.}] Let us consider the families $s=\left(e_\infty, e_c\right)$, $c\in\Gamma$, where $e_c\in E_c$, $e_\infty\in E_\infty$ and at least one of these numbers is odd. Let
\begin{equation}\label{34}
d=\frac{1}{2}\left(e_{\infty}-\sum\limits_{c\in\Gamma} e_c\right).
\end{equation}

If $d$ is a non -- negative integer, the family should be retained, otherwise it should be discarded.

\item[\underline{Step 3.}] For each family retained from Step~2, we form the rational function
\begin{equation}\label{35}
\theta=\frac{1}{2}\sum\limits_{c\in\Gamma}\frac{e_c}{x-c}
\end{equation}
and search for a monic polynomial $P$ of degree $d$ (the constant $d$ is defined by the formula~\eqref{52}) such, that
\begin{equation}\label{36}
P'''+3\theta P''+\left(3\theta^2+3\theta'-4R\right)P'+\left(\theta''+3\theta\theta'+\theta^3-4R\theta-2R'\right)P=0.
\end{equation}

If success is achieved, we set
\begin{equation*}
\varphi=\theta+\frac{P'}{P}
\end{equation*}
and let $\omega$ be a solution of the quadratic equation (algebraic equation of degree 2) of the form:
\begin{equation*}
\omega^2-\varphi\omega+\frac{1}{2}\varphi'+\frac{1}{2}\varphi^2-R=0.
\end{equation*}

Then
\begin{equation*}
\eta=\exp\left(\int\omega\left(x\right)dx\right) -
\end{equation*}
is a solution of the differential equation~\eqref{30}. If success is not achieved, then Case~2 cannot hold for the differential equation~\eqref{30}.
\end{description}

Similarly the Kovacic algorithm is stated to search for a liouvillian solutions of type 3 of the differential equation~\eqref{30}. Let us apply now this algorithm to search liouvillian solutions of the second order linear differential equation~\eqref{27}.

\subsection{Application of the Kovacic algorithm to the differential equation~\eqref{27}. General case.}

So, the differential equation being investigated has the form~\eqref{27}. In this equation we make a substitution according to~\eqref{29} and reduce it to the form~\eqref{30}:
\begin{equation}\label{37}
\frac{d^2y}{dx^2}=R\left(x\right)y.
\end{equation}

Here the function $R\left(x\right)$ takes the form:
\begin{equation}\label{38}
R\left(x\right)=\frac{U\left(x\right)}{V\left(x\right)},
\end{equation}
\begin{equation*}
\begin{array}{l}
U\left(x\right)=u_0x^{16}+u_1x^{15}+u_2x^{14}+u_3x^{13}+u_4x^{12}+u_5x^{11}+u_6x^{10}+u_7x^{9}+u_8x^{8}+\\ \\
+u_9x^{7}+u_{10}x^{6}+u_{11}x^{5}+u_{12}x^{4}+u_{13}x^{3}+u_{14}x^{2}+u_{15}x+u_{16},
\end{array}
\end{equation*}

\begin{equation*}
\begin{array}{l}
u_0=-4d_1^2\left(1+4d_1^2\right),\\ \\
u_1=-8d_1\left(1+2d_1^2\right)\left(B-2A\right),\\ \\
u_2=-8d_1^2\left(2d_1^2-1\right)\left(B^2-4h\right),\\ \\
u_3=16d_1\left(2d_1^2+5\right)\left(p_1-Bh\right)-4d_1\left(B-2A\right)\left(1+4d_1^2\right)\left(B^2-4h\right)+
4d_1\left(B-2A\right)^3,\\ \\
u_4=32d_1^2\left(7-2d_1^2\right)\left(B\left(p_1-Bh\right)-1+h^2\right)+2\left(B-2A\right)^2\left(B^2-4h\right)+
8\left(B-2A\right)\left(p_1-Bh\right)+\\ \\
+\left(B-2A\right)^4,\\ \\
u_5=32d_1^3\left(B^2-4h\right)\left(p_1-Bh\right)-64d_1^3B\left(B-2A\right)\left(p_1-Bh\right)+
152d_1\left(B^2-4h\right)\left(p_1-Bh\right)+\\ \\
+8d_1\left(6A+17B\right)\left(B-2A\right)\left(p_1-Bh\right)-
8d_1\left(B^2-4h\right)^2\left(B-2A\right)+4d_1\left(B^2-4h\right)
\left(B-2A\right)^3-\\ \\
-32d_1\left(B-2A\right)\left(5-2d_1^2\right)\left(1-h^2\right),\\ \\
u_6=32\left(1+17d_1^2-2d_1^4\right)\left(p_1-Bh\right)^2+96d_1^2B\left(B^2-4h\right)\left(p_1-Bh\right)+\\ \\
+16\left(B^2-4h\right)\left(B-2A\right)\left(p_1-Bh\right)+8\left(B-2A\right)^2\left(5B+2A\right)\left(p_1-Bh\right)-\\ \\
-32\left(\left(B-2A\right)^2+3d_1^2\left(B^2-4h\right)\right)\left(1-h^2\right)+4h\left(B-2A\right)^2\left(B^2+4AB-4A^2\right)-
\\ \\
-4\left(B-2A\right)^2\left(4+AB^2\left(B-A\right)\right),\\ \\
u_7=64d_1\left(\left(5+2d_1^2\right)B-\left(d_1^2-8\right)\left(B-2A\right)\right)\left(p_1-Bh\right)^2-
64d_1\left(5+2d_1^2\right)\left(1-h^2\right)\left(p_1-Bh\right)+\\ \\
+16Bd_1\left(B-2A\right)\left(\left(B-2A\right)^2+B^2-4h\right)\left(p_1-Bh\right)-\\ \\
-24d_1\left(B^2-4h\right)\left(\left(B-2A\right)^2-4\left(B^2-4h\right)\right)\left(p_1-Bh\right)-\\ \\
-16d_1\left(1-h^2\right)\left(B-2A\right)\left(\left(B-2A\right)^2+B^2-4h\right),\\ \\
u_8=8\left(24B^2d_1^2+\left(40d_1^2+7\right)\left(B^2-4h\right)+2\left(B-2A\right)\left(5B-18A\right)\right)\left(p_1-Bh\right)^2
+\\ \\
+8\left(B-2A\right)\left(3B^4-AB^3+2A^2B^2-4A^3B+40\right)\left(p_1-Bh\right)-\\ \\
-128\left(3Bd_1^2+2\left(B-2A\right)\right)\left(1-h^2\right)\left(p_1-Bh\right)-\\ \\
-32h\left(B-2A\right)\left(5B^2+2AB-4A^2\right)\left(p_1-Bh\right)+192d_1^2\left(1-h^2\right)^2-\\ \\
-4\left(1-h^2\right)\left(B-2A\right)^2\left(\left(B-2A\right)^2+2\left(B^2-4h\right)\right),\\ \\
\end{array}
\end{equation*}

\begin{equation*}
\begin{array}{l}
u_9=-128d_1\left(2-d_1^2\right)\left(p_1-Bh\right)^3+672Bd_1\left(B^2-4h\right)\left(p_1-Bh\right)^2-\\ \\
-1088d_1h\left(B-2A\right)\left(p_1-Bh\right)^2+64d_1\left(B-2A\right)\left(A^2+2AB+3B^2\right)\left(p_1-Bh\right)^2+\\ \\
+96d_1\left(1-h^2\right)\left(\left(B-2A\right)^2-7\left(B^2-4h\right)\right)\left(p_1-Bh\right),\\ \\
u_{10}=-32\left(B-22A-28Bd_1^2\right)\left(p_1-Bh\right)^3+12\left(B^2-4h\right)^2\left(p_1-Bh\right)^2-\\ \\
-64\left(5+14d_1^2\right)\left(1-h^2\right)\left(p_1-Bh\right)^2+80\left(B^2-4h\right)\left(B-2A\right)^2\left(p_1-Bh\right)^2+\\ \\
+128B\left(B-2A\right)\left(B^2-4h\right)\left(p_1-Bh\right)^2-4\left(B-2A\right)^3\left(7B+2A\right)\left(p_1-Bh\right)^2-\\ \\
-32\left(1-h^2\right)\left(B-2A\right)\left(4\left(B^2-4h\right)-\left(B-2A\right)^2\right)\left(p_1-Bh\right),\\ \\
u_{11}=32d_1\left(48B^2+9\left(B^2-4h\right)+\left(B-2A\right)\left(17B+6A\right)\right)\left(p_1-Bh\right)^3-\\ \\
-128d_1\left(1-h^2\right)\left(29B-10A\right)\left(p_1-Bh\right)^2+1536d_1\left(1-h^2\right)^2\left(p_1-Bh\right),\\ \\
u_{12}=16\left(35+32d_1^2\right)\left(p_1-Bh\right)^4+32\left(B+4A\right)\left(B^2-4h\right)\left(p_1-Bh\right)^3-\\ \\
-32\left(B-2A\right)^3\left(p_1-Bh\right)^3+192B\left(B-2A\right)\left(3B-2A\right)\left(p_1-Bh\right)^3-\\ \\ -96\left(1-h^2\right)\left(B^2-4h+2\left(B-2A\right)\left(5B-2A\right)\right)\left(p_1-Bh\right)^2\!+\!
384\left(B-2A\right)\left(1-h^2\right)^2\left(p_1-Bh\right),\\ \\
u_{13}=-64d_1\left(p_1-Bh\right)^3\left(36\left(1-h^2\right)+\left(14A-43B\right)\left(p_1-Bh\right)\right),\\ \\
u_{14}=-192\left(p_1-Bh\right)^3\left(\left(5h-\left(4B-3A\right)\left(B-A\right)\right)\left(p_1-Bh\right)+2\left(1-h^2\right)
\left(B-2A\right)\right),\\ \\
u_{15}=1152d_1\left(p_1-Bh\right)^5,\\ \\
u_{16}=-192\left(p_1-Bh\right)^4\left(1-h^2-2\left(B-A\right)\left(p_1-Bh\right)\right),
\end{array}
\end{equation*}
\begin{equation*}
V\left(x\right)=4\left(P_6\left(x\right)\right)^2\left(P_3\left(x\right)\right)^2.
\end{equation*}

Thus, it is easy to see, that the function $R\left(x\right)$ has 9 finite poles of the second order. We will assume that all these poles are distinct. This assumption take place for almost all values of parameters of the problem. Let us denote the roots of the polynomial $P_6\left(x\right)$ by $x_1$, $x_2$, $x_3$, $x_4$, $x_5$, $x_6$. Note that this polynomial contains only the terms of even degree, therefore its roots satisfy the conditions:
\begin{equation*}
x_2=-x_1,\qquad x_4=-x_3,\qquad x_6=-x_5.
\end{equation*}

Let us denote the roots of the polynomial $P_3\left(x\right)$ by $x_7$, $x_8$, $x_9$. Now let us consider the partial fraction expansion of the function $R\left(x\right)$. It has the form:
\begin{equation*}
R\left(x\right)=-\frac{3}{16}\sum\limits_{i=1}^6\frac{1}{\left(x-x_i\right)^2}+\sum\limits_{i=1}^9\frac{\gamma_i\left(x_i\right)}
{x-x_i}+\frac{3}{4}\sum\limits_{i=7}^9\frac{1}{\left(x-x_i\right)^2}.
\end{equation*}

The coefficients $\gamma_i\left(x_i\right)$, $i=1, 2,\ldots 9$ have a very complicated form and we do not write them explicitly here. It is possible to note the following properties of the partial fraction expansion of the function $R\left(x\right)$.
\begin{enumerate}
\item The coefficients $b_1,\ldots, b_6$ of $\displaystyle\frac{b_i}{\left(x-x_i\right)^2}$, $i=1,\ldots, 6$ are all equal
\begin{equation*}
b_i=-\frac{3}{16}, \quad i=1,\ldots, 6.
\end{equation*}
\item The coefficients $b_7$, $b_8$, $b_9$ of $\displaystyle\frac{b_i}{\left(x-x_i\right)^2}$, $i=7, 8, 9$ are all equal
\begin{equation*}
b_i=\frac{3}{4}, \quad i=7, 8, 9.
\end{equation*}
\item The Laurent expansion of $R\left(x\right)$ at $x=\infty$ has the form:
\begin{equation*}
R\left(x\right)=-\frac{\left(1+4d_1^2\right)}{4x^2}+O\left(\frac{1}{x^4}\right)
\end{equation*}
\end{enumerate}

Thus, we have
\begin{equation*}
b_{\infty}=-\frac{1}{4}-d_1^2,
\end{equation*}
and therefore
\begin{equation*}
1+4b_{\infty}=-4d_1^2
\end{equation*}

This means that the numbers $\alpha^{\pm}_{\infty}$ calculating during the application of the Kovacic algorithm for searching the liouvillian solutions of type 1, are complex numbers if $d_1\ne 0$. All the remaining numbers $\alpha^{\pm}_c$ are rational. They are presented in the following Table. Therefore, the number $d$, calculated by formula~\eqref{31} in the process of searching for liouvillian solutions of type 1, is a complex number for $d_1\ne 0$. This fact indicates the absence of liouvillian solutions of type 1 for $d_1\ne 0$.
\begin{center}
\begin{tabular}{ |c|c|c|c|c|c|c|c|c|c| }
 \hline
  & $x_1$ & $x_2$ & $x_3$ & $x_4$ & $x_5$ & $x_6$ & $x_7$ & $x_8$ & $x_9$ \\ \hline
  \rule{0pt}{4ex} $\alpha_c^+$ & $\displaystyle\frac{3}{4}$ & $\displaystyle\frac{3}{4}$ & $\displaystyle\frac{3}{4}$ & $\displaystyle\frac{3}{4}$ & $\displaystyle\frac{3}{4}$ & $\displaystyle\frac{3}{4}$ & $\displaystyle\frac{3}{2}$ & $\displaystyle\frac{3}{2}$ & $\displaystyle\frac{3}{2}$\\[2ex] \hline
 \rule{0pt}{4ex} $\alpha_c^-$ & $\displaystyle\frac{1}{4}$ & $\displaystyle\frac{1}{4}$ & $\displaystyle\frac{1}{4}$ & $\displaystyle\frac{1}{4}$ & $\displaystyle\frac{1}{4}$ & $\displaystyle\frac{1}{4}$ & $-\displaystyle\frac{1}{2}$ & $-\displaystyle\frac{1}{2}$ & $-\displaystyle\frac{1}{2}$\\[2ex] \hline
\end{tabular}
{\vskip 0.2cm}
Table. Numbers $\alpha^{\pm}_c$ for searching liouvillian solutions of type 1.
\end{center}

Moreover, the coefficient $b_{\infty}$ coincides with the number $\gamma$ calculating during the checking of the necessary conditions of existence of liouvillian solutions of type 3 for the differential equation~\eqref{37}. According to this necessary conditions for existence of liouvillian solutions of type 3 the number
\begin{equation*}
\sqrt{1+4\gamma}=\sqrt{1+4b_{\infty}}
\end{equation*}
should be rational. However, when $d_1\ne 0$ this number is pure imaginary. Thus, we can state that for $d_1\ne 0$ the second order linear differential equation~\eqref{37} (or~\eqref{27}) do not have liouvillian solutions of type 3. Thus, the following Theorem is valid.
\begin{thm}\label{Theorem4}
If all roots of polynomials $P_3\left(x\right)$ and $P_6\left(x\right)$ are distinct and $d_1\ne 0$, then the problem of motion of a heavy gyrostat with a fixed point in the Hess -- Sretensky case has not liouvillian solutions of type $1$ and type $3$. $\Box$
\end{thm}

So we are going to study the problem of existence of liouvillian solutions of type 2 for the differential equation~\eqref{37} (or~\eqref{27}). The necessary conditions for the existence of such a solution are fulfilled (see~\ref{Theorem3}). We apply step by step the Kovacic algorithm as it was described in Section 5.

\underline{{\bf Step 1.}} According to the Kovacic algorithm we firstly define the sets $E_c$ and $E_{\infty}$ for every pole of the function $R\left(x\right)$. For the finite poles $x=x_i$, $i=1,\ldots, 6$, which are roots of the polynomial $P_6\left(x\right)$, these sets $E_{x_i}$ have the form:
\begin{equation*}
E_{x_i}=\{1, 2, 3\}, \quad i=1,\ldots, 6.
\end{equation*}

For the finite poles $x=x_i$, $i=7, 8, 9$, which are roots of the polynomial $P_3\left(x\right)$, these sets $E_{x_i}$ have the form:
\begin{equation*}
E_{x_i}=\{-2, 2, 6\}, \quad i=7, 8, 9.
\end{equation*}

The set $E_{\infty}$ contains only one element and this set equals
\begin{equation*}
E_{\infty}=\{2\}.
\end{equation*}

\underline{{\bf Step 2.}} Now we consider all possible tuples
\begin{equation*}
s=\left(e_{\infty}, e_{x_1}, e_{x_2}, e_{x_3}, e_{x_4}, e_{x_5}, e_{x_6}, e_{x_7}, e_{x_8}, e_{x_9}\right)
\end{equation*}
of elements of the sets $E_{\infty}$, $E_{x_i}$, $i=1,\ldots, 9$; at least one of the elements in each tuple must be odd. Using~\eqref{34}, for each tuple $s$ we get
\begin{equation*}
d=\frac{1}{2}\left(e_{\infty}-e_{x_1}-e_{x_2}-e_{x_3}-e_{x_4}-e_{x_5}-e_{x_6}-e_{x_7}-e_{x_8}-e_{x_9}\right).
\end{equation*}

According to the algorithm, the constant $d$ must be nonnegative integer. Note that the minimal value of the sum of the elements of sets corresponding to finite poles is zero. Therefore the maximal value of $d$, calculated according to~\eqref{34}, equals $d=1$. The value $d=1$ corresponds to the set $s_1$, in which the elements $e_{\infty} $ and $e_{x_i}$, $i=1, 2, \ldots 9$ are equal
\begin{equation*}
s_1=\left(2, 1, 1, 1, 1, 1, 1, -2, -2, -2\right).
\end{equation*}

We also have several tuples for which the constant $d$, calculated by the formula~\eqref{34}, equals to zero. The following tuples correspond to $d=0$:
\begin{equation*}
\begin{array}{l}
s_2=\left(2, 3, 1, 1, 1, 1, 1, -2, -2, -2\right),\quad s_3=\left(2, 1, 3, 1, 1, 1, 1, -2, -2, -2\right),\\ \\
s_4=\left(2, 1, 1, 3, 1, 1, 1, -2, -2, -2\right),\quad s_5=\left(2, 1, 1, 1, 3, 1, 1, -2, -2, -2\right),\\ \\
s_6=\left(2, 1, 1, 1, 1, 3, 1, -2, -2, -2\right),\quad s_7=\left(2, 1, 1, 1, 1, 1, 3, -2, -2, -2\right),\\ \\
s_8=\left(2, 2, 2, 1, 1, 1, 1, -2, -2, -2\right),\quad s_9=\left(2, 2, 1, 2, 1, 1, 1, -2, -2, -2\right),\\ \\
s_{10}=\left(2, 2, 1, 1, 2, 1, 1, -2, -2, -2\right),\quad s_{11}=\left(2, 2, 1, 1, 1, 2, 1, -2, -2, -2\right),
\end{array}
\end{equation*}
\begin{equation*}
\begin{array}{l}
s_{12}=\left(2, 2, 1, 1, 1, 1, 2, -2, -2, -2\right),\quad s_{13}=\left(2, 1, 2, 2, 1, 1, 1, -2, -2, -2\right),\\ \\
s_{14}=\left(2, 1, 2, 1, 2, 1, 1, -2, -2, -2\right),\quad s_{15}=\left(2, 1, 2, 1, 1, 2, 1, -2, -2, -2\right),\\ \\
s_{16}=\left(2, 1, 2, 1, 1, 1, 2, -2, -2, -2\right),\quad s_{17}=\left(2, 1, 1, 2, 2, 1, 1, -2, -2, -2\right),\\ \\
s_{18}=\left(2, 1, 1, 2, 1, 2, 1, -2, -2, -2\right),\quad s_{19}=\left(2, 1, 1, 2, 1, 1, 2, -2, -2, -2\right),\\ \\
s_{20}=\left(2, 1, 1, 1, 2, 2, 1, -2, -2, -2\right),\quad s_{21}=\left(2, 1, 1, 1, 2, 1, 2, -2, -2, -2\right),\\ \\
s_{22}=\left(2, 1, 1, 1, 1, 2, 2, -2, -2, -2\right).
\end{array}
\end{equation*}

We must check all these tuples.

\underline{{\bf Step 3.}} We start with the tuple $s_1$. According to the algorithm, let us find the function $\theta$ by the formula~\eqref{35}. Since the elements at all finite poles $x=x_i$, $i=1, 2, \ldots, 6$ are the same, and the elements at all poles $x=x_i$, $i=7, 8, 9$ are the same, then we can write the function $\theta$ in explicit form. For the set $s_1$ this function has the form
\begin{equation*}
\theta=\frac{3x^5+2\left(B^2-4h\right)x^3+4\left(\left(p_1-Bh\right)B-1+h^2\right)x}{P_6\left(x\right)}-
\frac{2\left(3d_1x^2+\left(B-2A\right)x\right)}{P_3\left(x\right)}.
\end{equation*}

The polynomial $P$ of degree $d=1$
\begin{equation*}
P=x+K_1
\end{equation*}
should identically satisfy the differential equation~\eqref{36}. After substitution of $P=x+K_1$ and the functions $\theta$ and $R\left(x\right)$ to the equation~\eqref{36}, we obtain in the left hand side of~\eqref{36} the rational expression. The numerator of this expression has a form of the ninth degree polynomial:
\begin{equation*}
\Pi_9=-4d_1^2\left(K_1\left(4d_1^2+1\right)-2d_1\left(B-2A\right)\right)x^9+\cdots
\end{equation*}

Let us set
\begin{equation*}
K_1=\frac{2d_1\left(B-2A\right)}{1+4d_1^2}.
\end{equation*} 

Then the numerator of the rational expression in the left hand side of~\eqref{36} takes the form:
\begin{equation*}
\Pi_8=\frac{16d_1^2\left(1+d_1^2\right)\left(B-2A\right)^2}{1+4d_1^2}x^8+\cdots
\end{equation*}

To turn the leading coefficient of this polynomial to zero, we set
\begin{equation*}
A=\frac{B}{2}.
\end{equation*}

Then the numerator of the rational expression in the left hand side of~\eqref{36} takes the form:
\begin{equation*}
\Pi_7=-48d_1^3\left(p_1-Bh\right)x^7+\cdots
\end{equation*}

If we set
\begin{equation*}
p_1=Bh,
\end{equation*} 
then this polynomial $\Pi_7$ becomes zero. Therefore we can state the following Theorem based on the verification of the tuple $s_1$.

\begin{thm}\label{Theorem5}
For the existence of liouvillian solutions of type 2 in the problem of motion of a heavy gyrostat with a fixed point in the Hess -- Sretensky case the following conditions must be satisfied:
\begin{equation}\label{39}
p_1=Bh,\quad B=2A.\quad \Box
\end{equation}
\end{thm}

Similarly we can check the other tuples $s_2, \ldots, s_{22}$. Checking of these tuples gives us the same conditions as checking of the tuple $s_1$. Finally we can conclude that conditions~\eqref{39} are necessary and sufficient conditions of existence of liouvillian solutions of the second order linear differential equation~\eqref{37} (or~\eqref{27}).

Under conditions~\eqref{39}, the coefficients $a\left(x\right)$ and $b\left(x\right)$ of the differential equation~\eqref{27} take very simple form:
\begin{equation*}
a\left(x\right)=\frac{x^4-4h^2+4}{\left(x^4+4\left(A^2-h\right)x^2+4h^2-4\right)x},\; b\left(x\right)=\frac{d_1^2x^2}{x^4+4\left(A^2-h\right)x^2+4h^2-4},
\end{equation*}
and the general solution of the linear differential equation~\eqref{27} is written in the following closed form:
\begin{equation*}
u\left(x\right)=C_1\cos\Phi\left(x\right)+C_2\sin\Phi\left(x\right),
\end{equation*}
\begin{equation*}
\Phi\left(x\right)=\int\frac{d_1x}{\sqrt{x^4+4\left(A^2-h\right)x^2+4h^2-4}}dx,
\end{equation*}
where $C_1$ and $C_2$ are arbitrary constants.

The first of conditions~\eqref{39} means that the constant of the reduced first integral
\begin{equation*}
\frac{B\left(y^2+z^2\right)}{2}-y\nu_2-z\nu_3=Bh-p_1
\end{equation*}
should be zero. This first integral can be expressed through the initial parameters and variables as follows:
\begin{equation*}
\begin{array}{l}
\displaystyle\frac{\left(A_1x_1s_2-A_2x_2s_1\right)}{2\left(A_1-A_2\right)x_1x_2}\left[
\displaystyle\frac{\left(\left(A_2\omega_2+s_2\right)x_1-\left(A_1\omega_1+s_1\right)x_2\right)^2}
{A_3\left(x_1^2+x_2^2\right)}+A_3\omega_3^2\right]-\\ \\
-\displaystyle\frac{Mg}{x_1^2+x_2^2}\left(\left(A_2\omega_2+s_2\right)x_1-\left(A_1\omega_1+s_1\right)x_2\right)\left(\gamma_2x_1-
\gamma_1x_2\right)-MgA_3\omega_3\gamma_3=0.
\end{array}
\end{equation*}

The second of conditions~\eqref{39} in the initial parameters can be written in the following symmetrical form:
\begin{equation}\label{40}
A_1x_1\left(A_2\left(x_1^2+x_2^2\right)+\left(A_2-A_1\right)x_1^2\right)s_2\!=\!
A_2x_2\left(A_1\left(x_1^2+x_2^2\right)+\left(A_1-A_2\right)x_2^2\right)s_1.
\end{equation}
or in the form
\begin{equation}\label{41}
\left(2A_3-A_1\right)x_1s_2=\left(2A_3-A_2\right)x_2s_1.
\end{equation}

The right-hand side of condition~\eqref{40} (or~\eqref{41}) can be obtained from the left-hand side by a cyclic permutation of the indices 1 and 2 in the left-hand side.

Thus, we proved that the solution of the problem of motion of a heavy gyrostat with a fixed point is reduced to solving the second order linear homogeneous differential equation with rational coefficients. By application of the Kovacic algorithm to this linear differential equation, we found the conditions of existence of liouvillian solutions of the corresponding linear differential equation. Under these conditions the equations of motion of the gyrostat can be integrated in quadratures.  

\subsection{Motion of a gyrostat under the action of additional gyroscopic forces.}

Let us consider now the more general problem when the gyrostat moves under the action of the moment of gravity
\begin{equation*}
Mg\left[\vec{\gamma}\times\vec{r}_0\right]
\end{equation*}
and also under the action of the moment of gyroscopic forces
\begin{equation*}
\left[\mathbb{G}\vec{\gamma}\times\vec{\omega}\right].
\end{equation*}

Here $\mathbb{G}=\mathbb{G}^{\tt T}$ is a symmetric matrix with constant components. It is worth to note that a moment of this type arises due to the action of the Lorentz force on a rotating electrically charged body in a uniform magnetic field with a constant field vector (see,~\cite{Lunev,Kozlov,Samsonov,Kosov}). In this case equations of motion of a gyrostat has the form
\begin{equation}\label{42}
\mathbb{I}\dot{{\vec\omega}}+\left[{\vec\omega}\times\left(\mathbb{I}{\vec\omega}+{\vec s}\right)\right]=Mg\left[\boldsymbol{\gamma}\times\boldsymbol{r}_0\right]+
\left[\mathbb{G}\boldsymbol{\gamma}\times\boldsymbol{\omega}\right],\qquad \dot{\vec{\gamma}}=\left[\vec{\gamma}\times\vec{\omega}\right].
\end{equation}

For any values of parameters of the system and for any initial conditions the system~\eqref{42} possesses three first integrals --- the energy integral
\begin{equation*}
H=\displaystyle\frac{1}{2}\left(\vec{\omega}\cdot\mathbb{I}\vec{\omega}\right)+Mg\left(\vec{\gamma}\cdot
\vec{r}_0\right)=E,
\end{equation*}
the area integral
\begin{equation*}
K=\left(\left(\mathbb{I}\vec{\omega}+\vec{s}\right)\cdot\vec{\gamma}\right)+\frac{1}{2}\left(\mathbb{G}
\vec{\gamma}\cdot\vec{\gamma}\right)=k
\end{equation*}
and the geometrical integral
\begin{equation*}
\left(\vec{\gamma}\cdot\vec{\gamma}\right)=1.
\end{equation*}

A.~A.~Kosov in his paper~\cite{Kosov} obtained the conditions under which there exists a special integrable case of the system of equations~\eqref{42}, similar to the Hess -- Sretensky case. The following Theorem holds.

\begin{thm}\label{Theorem6}
Let us assume that for the system of equations~\eqref{42} the following conditions are true:
\begin{enumerate}
\item
\begin{equation}\label{43}
A_2\left(A_3-A_1\right)x_2^2=A_1\left(A_2-A_3\right)x_1^2,\quad x_3=0, \quad s_3=0, \quad A_2>A_3>A_1;
\end{equation}

\item The matrix $\mathbb{G}$ has the form:
\begin{equation*}
\mathbb{G}=\left(\begin{array}{ccc}
g_{11}& g_{12}& 0\\
g_{12} & g_{22}& 0 \\
0& 0& 0
\end{array}
\right),
\end{equation*}
and its components $g_{11}$, $g_{12}$, $g_{22}$ satisfy the following conditions
\begin{equation}\label{44}
g_{11}x_2-g_{12}x_1=0,\quad g_{12}x_2-g_{22}x_1=0.
\end{equation}
\end{enumerate}

Therefore the system of equations~\eqref{42} admits the additional special Hess -- Sretensky integral of the form
\begin{equation}\label{45}
A_1\omega_1 x_1+A_2\omega_2 x_2+\frac{A_1x_1\left(s_2x_1-s_1x_2\right)}{\left(A_3-A_1\right)x_2}=0. \quad \Box
\end{equation}
\end{thm}

Let us study the conditions~\eqref{44}. We have two equations for three components of matrix $\mathbb{G}$. Therefore, using the auxiliary multiplier $\lambda$ we can rewrite them as follows
\begin{equation*}
g_{11}=\lambda x_1^2,\quad g_{12}=\lambda x_1x_2,\quad g_{22}=\lambda x_2^2.
\end{equation*}

Thus, if the conditions~\eqref{43},~\eqref{44} are met, then the system of equations~\eqref{42} in scalar form can be written as follows:
\begin{equation}\label{46}
\begin{array}{l}
A_1{\dot\omega}_1+\left(A_3-A_2\right)\omega_2\omega_3-s_2\omega_3=-Mgx_2\gamma_3+\lambda x_2\omega_3\left(\gamma_1x_1+\gamma_2x_2\right),\\ \\
A_2{\dot\omega}_2+\left(A_1-A_3\right)\omega_1\omega_3+s_1\omega_3=Mgx_1\gamma_3-\lambda x_1\omega_3\left(\gamma_1x_1+\gamma_2x_2\right),\\ \\
A_3{\dot\omega}_3+\left(A_2-A_1\right)\omega_1\omega_2+s_2\omega_1-s_1\omega_2=Mg\left(x_2\gamma_1-x_1\gamma_2\right)+
\lambda\left(x_1\omega_2-x_2\omega_1\right)\left(\gamma_1x_1+\gamma_2x_2\right);
\end{array}
\end{equation}
\begin{equation}\label{47}
\dot{\gamma}_1=\omega_3\gamma_2-\omega_2\gamma_3,\quad \dot{\gamma}_2=\omega_1\gamma_3-\omega_3\gamma_1,\quad
\dot{\gamma}_3=\omega_2\gamma_1-\omega_1\gamma_2.
\end{equation}

The system of equations~\eqref{46},~\eqref{47} possesses the first integrals
\begin{equation*}
\begin{array}{l}
\displaystyle\frac{1}{2}\left(A_1\omega_1^2+A_2\omega_2^2+A_3\omega_3^2\right)+Mg\left(x_1\gamma_1+x_2\gamma_2\right)=E,\quad
\gamma_1^2+\gamma_2^2+\gamma_3^2=1, \\ \\
\left(A_1\omega_1+s_1\right)\gamma_1+\left(A_2\omega_2+s_2\right)\gamma_2+A_3\omega_3\gamma_3+\displaystyle\frac{\lambda}{2}
\left(x_1\gamma_1+x_2\gamma_2\right)^2=k.
\end{array}
\end{equation*}

Let us prove that under conditions~\eqref{43},~\eqref{44} the system of equations~\eqref{46},~\eqref{47} admits the additional special integral~\eqref{45}. We multiply the first and second equations of the system~\eqref{46} by $x_1$ and $x_2$ respectively and add them together:
\begin{equation}\label{48}
\frac{d}{dt}\left(A_1\omega_1 x_1+A_2\omega_2 x_2\right)=\left(A_3-A_1\right)\omega_1\omega_3 x_2+\left(A_2-A_3\right)\omega_2\omega_3 x_1+\omega_3\left(s_2x_1-s_1x_2\right).
\end{equation}

Using the conditions~\eqref{43},~\eqref{44} the right hand side of the equation~\eqref{48} can be rewritten in the form
\begin{equation}\label{49}
\begin{array}{l}
\left(A_3-A_1\right)\omega_1\omega_3 x_2+\left(A_2-A_3\right)\omega_2\omega_3 x_1+\omega_3\left(s_2x_1-s_1x_2\right)=\\ \\
=\displaystyle\frac{\omega_3\left(A_2-A_3\right)x_1}{A_2x_2}\left(A_1\omega_1 x_1+A_2\omega_2 x_2
+\displaystyle\frac{A_1x_1\left(s_2x_1-s_1x_2\right)}{\left(A_3-A_1\right)x_2}\right).
\end{array}
\end{equation}

Therefore, taking into account~\eqref{49}, from the equation~\eqref{48} we can derive that if condition~\eqref{45} is satisfied at the initial moment of time, then this condition will hold during the whole time of motion of a gyrostat. Thus, we proved that under the conditions~\eqref{43},~\eqref{44} the system of equations~\eqref{46},~\eqref{47} admits the additional special integral~\eqref{45}.

\subsection{Equations of motion of a gyrostat in the special coordinate system.}

We shall use again the variables $L_1$, $L_2$, $L_3$, $\nu_1$, $\nu_2$, $\nu_3$ and parameters $a$, $b$, $c$, $k_1$, $k_2$, $\Gamma$, which have been introduced in Section 3. Using the variables $L_1$, $L_2$, $L_3$, $\nu_1$, $\nu_2$, $\nu_3$ we can rewrite the Euler -- Poisson equations~\eqref{46},~\eqref{47} as follows:
\begin{equation}\label{50}
\begin{array}{lll}
\dot{L}_1&=&-bL_3\left(L_1-\displaystyle\frac{ck_2}{b}\right),\quad
\dot{L}_2=\left(a-c\right)L_1L_3+bL_2L_3-ck_1L_3-\Lambda c\nu_1L_3+\Gamma\nu_3,
\\ \\
\dot{L}_3&=&-\left(a-c\right)L_1L_2+bL_1^2-bL_2^2+\left(k_1b-k_2a\right)L_1+\left(k_1c-k_2b\right)L_2+\\ \\
&+&\Lambda\left(bL_1+cL_2\right)\nu_1-\Gamma\nu_2,\\ \\
\dot{\nu}_1&=&cL_3\nu_2-\left(cL_2+bL_1\right)\nu_3,\quad \dot{\nu}_2=\left(aL_1+bL_2\right)\nu_3-cL_3\nu_1,\\ \\
\dot{\nu}_3&=&-\left(aL_1+bL_2\right)\nu_2+\left(cL_2+bL_1\right)\nu_1.
\end{array}
\end{equation}

Here $\Lambda$ is a parameter which characterizes the gyroscopic forces. It has the form
\begin{equation*}
\Lambda=\lambda\left(x_1^2+x_2^2\right).
\end{equation*}

The additional first integral~\eqref{45} that exists in the Hess -- Sretensky case can be found explicitly from the first equation of the system~\eqref{50}. It has the form:
\begin{equation}\label{51}
L_1\equiv\displaystyle\frac{ck_2}{b}.
\end{equation}

The invariant manifold~\eqref{51} (or~\eqref{45}, using different notations) together with conditions~\eqref{43},~\eqref{44} defines the Hess -- Sretensky integrable case of motion of a heavy gyrostat with a fixed point under the action of gyroscopic forces. Under conditions~\eqref{43},~\eqref{44} and taking into account the special first integral~\eqref{51}, equations~\eqref{50} are noticeably simplified and take the form:
\begin{equation}\label{52}
\begin{array}{l}
\dot{\tilde{L}}_2=b\tilde{L}_2L_3+\left(F-Gc\right)L_3-\Lambda c\nu_1L_3+\Gamma\nu_3,\quad \dot{L}_3=-b{\tilde L}_2^2-\left(F-Gc\right){\tilde L}_2+\Lambda c{\tilde L}_2\nu_1-\Gamma\nu_2,\\ \\
\dot{\nu}_1=cL_3\nu_2-c{\tilde L}_2\nu_3,\quad \dot{\nu}_2=-cL_3\nu_1+b{\tilde L}_2\nu_3+F\nu_3,\quad
\dot{\nu}_3=c{\tilde L}_2\nu_1-b{\tilde L}_2\nu_2-F\nu_2.
\end{array}
\end{equation}

Here we again introduce the following notations
\begin{equation*}
\tilde{L}_2=L_2+k_2,\quad F=\frac{\left(ac-b^2\right)k_2}{b},\quad G=\frac{ck_2}{b}+k_1.
\end{equation*}
 
Equations~\eqref{52} possess the following first integrals
\begin{equation}\label{53}
\begin{array}{l}
\displaystyle\frac{c}{2}\left({\tilde L}_2^2+L_3^2\right)+\Gamma\nu_1=E;\quad {\tilde L}_2\nu_2+L_3\nu_3+G\nu_1+\frac{\Lambda}{2}\nu_1^2=k;\quad
\nu_1^2+\nu_2^2+\nu_3^2=1.
\end{array}
\end{equation}

\subsection{Equations of Motion in Dimensionless Form. Derivation of the Second-order Differential Equation.}

Now let us write the equations~\eqref{52} and its first integrals~\eqref{53} in dimensionless form. For this purpose we introduce the dimensionless variables $y$ and $z$ using formulas
\begin{equation*}
y={\tilde L}_2\sqrt{\displaystyle\frac{c}{\Gamma}},\quad z=L_3\sqrt{\displaystyle\frac{c}{\Gamma}},
\end{equation*}
the dimensionless time $\tau=t\sqrt{\Gamma c}$ and the dimensionless parameters
\begin{equation*}
d_1=\displaystyle\frac{b}{c},\quad Q=\Lambda\sqrt{\displaystyle\frac{c}{\Gamma}},\quad A=\displaystyle\frac{F}{\sqrt{\Gamma c}},\quad B=G\sqrt{\displaystyle\frac{c}{\Gamma}}.
\end{equation*}

We also introduce the dimensionless constants of the first integrals
\begin{equation*}
h=\displaystyle\frac{E}{\Gamma},\quad p_1=k\sqrt{\displaystyle\frac{c}{\Gamma}}.
\end{equation*}

Now we can rewrite the system of equations of a gyrostat in dimensionless form:
\begin{equation}\label{54}
\begin{array}{l}
\displaystyle\frac{dy}{d\tau}=d_1yz+\left(A-B\right)z-Q\nu_1z+\nu_3,\quad \displaystyle\frac{dz}{d\tau}=-d_1y^2-\left(A-B\right)y+Q\nu_1y-\nu_2,\\ \\
\displaystyle\frac{d{\nu}_1}{d\tau}=z\nu_2-y\nu_3,\quad \displaystyle\frac{d{\nu}_2}{d\tau}=d_1y\nu_3-z\nu_1+A\nu_3,\quad
\displaystyle\frac{d{\nu}_3}{d\tau}=-d_1y\nu_2+y\nu_1-A\nu_2,\\ \\
\end{array}
\end{equation}

The system of equations~\eqref{54} possesses three first integrals
\begin{equation}\label{55}
\displaystyle\frac{y^2+z^2}{2}+\nu_1=h,\quad y\nu_2+z\nu_3+B\nu_1+\frac{Q}{2}\nu_1^2=p_1,\quad \nu_1^2+\nu_2^2+\nu_3^2=1.
\end{equation}

From the system~\eqref{54}, using the first integrals~\eqref{55}, we will obtain now the second-order linear differential equation. Multiplying the first equation of the system~\eqref{54} by $y$ and the second by $z$ and adding them, we get:
\begin{equation}\label{56}
\frac{d}{d\tau}\left(\frac{y^2+z^2}{2}\right)=y\nu_3-z\nu_2.
\end{equation}

From the first equation of the system~\eqref{55} we can find that
\begin{equation*}
\nu_1=h-\displaystyle\frac{y^2+z^2}{2},
\end{equation*}
and, therefore, we have:
\begin{equation*}
\begin{array}{l}
\nu_2^2+\nu_3^2=1-\nu_1^2=1-\left(\displaystyle\frac{y^2+z^2}{2}-h\right)^2,\\ \\
y\nu_2+z\nu_3=p_1+B\left(\displaystyle\frac{y^2+z^2}{2}-h\right)-\displaystyle\frac{Q}{2}
\left(\displaystyle\frac{y^2+z^2}{2}-h\right)^2.
\end{array}
\end{equation*}

Therefore, from the trivial identity
\begin{equation}\label{57}
\left(y^2+z^2\right)\left(\nu_2^2+\nu_3^2\right)=\left(y\nu_2+z\nu_3\right)^2+\left(y\nu_3-z\nu_2\right)^2,
\end{equation}
we obtain:
\begin{equation*}
\begin{array}{l}
\left(y\nu_3-z\nu_2\right)^2=\left(y^2+z^2\right)\left(1-\left(\displaystyle\frac{y^2+z^2}{2}-h\right)^2\right)-\\ \\
-\left(p_1+B\left(\displaystyle\frac{y^2+z^2}{2}-h\right)-\displaystyle\frac{Q}{2}
\left(\displaystyle\frac{y^2+z^2}{2}-h\right)^2\right)^2=\displaystyle\frac{P_4\left(y^2+z^2\right)}{64},
\end{array}
\end{equation*}
where $P_4\left(y^2+z^2\right)$ is a fourth-degree polynomial of variable $y^2+z^2$ with constant coefficients. In explicit form the polynomial $P_4\left(y^2+z^2\right)$ can be written as follows:
\begin{equation*}
\begin{array}{l}
P_4\left(y^2+z^2\right)=-Q^2\left(y^2+z^2\right)^4+8\left(BQ+Q^2h-2\right)\left(y^2+z^2\right)^3+\\ \\
+8\left(8h-2B^2-6BQh-3Q^2h^2+2p_1Q\right)\left(y^2+z^2\right)^2-16\left(Qh^2+2Bh-2p_1\right)^2+\\ \\
+32\left(2-2h^2-2p_1B-2p_1Qh+2B^2h+3QBh^2+Q^2h^3\right)\left(y^2+z^2\right).
\end{array}
\end{equation*}

We will take that
\begin{equation}\label{58}
y\nu_3-z\nu_2=-\displaystyle\frac{\sqrt{P_4\left(y^2+z^2\right)}}{8},
\end{equation}
(we can choose any of the sign before the square root in~\eqref{58}). Taking into account~\eqref{58} we can rewrite~\eqref{56} as follows:
\begin{equation*}
\frac{d}{d\tau}\left(\frac{y^2+z^2}{2}\right)=-\displaystyle\frac{\sqrt{P_4\left(y^2+z^2\right)}}{8}.
\end{equation*}

Now we multiply the second equation of the system~\eqref{54} by $y$, the first equation of the system~\eqref{54} by $-z$ and add them. As a result, we get:
\begin{equation*}
\begin{array}{l}
y\displaystyle\frac{dz}{d\tau}-z\displaystyle\frac{dy}{d\tau}=-d_1y\left(y^2+z^2\right)-\left(A-B\right)\left(y^2+z^2\right)+
Q\nu_1\left(y^2+z^2\right)-\left(y\nu_2+z\nu_3\right)=\\ \\
=-d_1y\left(y^2+z^2\right)-\displaystyle\frac{3Q}{8}\left(y^2+z^2\right)^2+\displaystyle\frac{1}{2}\left(Qh+B-2A\right)
\left(y^2+z^2\right)+\displaystyle\frac{1}{2}\left(Qh^2+2Bh-2p_1\right).
\end{array}
\end{equation*}

We introduce now new variables $x$ and $\varphi$ ("polar coordinates") by putting:
\begin{equation}\label{59}
y=x\cos\varphi,\quad z=x\sin\varphi.
\end{equation}

Then for the variables $x$ and $\varphi$ we have the following system of two differential equations:
\begin{equation}\label{60}
\begin{array}{l}
x\displaystyle\frac{dx}{d\tau}=-\displaystyle\frac{\sqrt{P_8\left(x\right)}}{8},\\ \\
x^2\displaystyle\frac{d\varphi}{d\tau}=-d_1x^3\cos\varphi-\displaystyle\frac{3Q}{8}x^4+\displaystyle\frac{1}{2}
\left(Qh+B-2A\right)x^2-\displaystyle\frac{1}{2}\left(Qh^2+2Bh-2p_1\right),
\end{array}
\end{equation}
\begin{equation*}
\begin{array}{l}
P_8\left(x\right)=-Q^2x^8+8\left(BQ+Q^2h-2\right)x^6+8\left(8h-2B^2-6BQh-3Q^2h^2+2p_1Q\right)x^4+\\ \\
+32\left(2-2h^2-2p_1B-2p_1Qh+2B^2h+3QBh^2+Q^2h^3\right)x^2-16\left(Qh^2+2Bh-2p_1\right)^2.
\end{array}
\end{equation*}

From the system of equations~\eqref{60}, dividing one equation by the other, we obtain the differential equation for function $\varphi=\varphi\left(x\right)$:
\begin{equation}\label{61}
\frac{d\varphi}{dx}=\frac{8d_1x^2}{\sqrt{P_8\left(x\right)}}\cos\varphi+
\displaystyle\frac{3Qx^4-4\left(Qh+B-2A\right)x^2+4\left(2p_1-Qh^2-2Bh\right)}{x\sqrt{P_8\left(x\right)}}.
\end{equation}

The equation~\eqref{61} has the form
\begin{equation*}
\frac{d\varphi}{dx}=a_0\cos\varphi+b_0,
\end{equation*}
therefore, using the following change of variable
\begin{equation*}
w=\tan\frac{\varphi}{2},
\end{equation*}
it can be reduced to the Riccati equation:
\begin{equation}\label{62}
\frac{dw}{dx}=\frac{b_0-a_0}{2}w^2+\frac{b_0+a_0}{2}=f_2w^2+f_0,
\end{equation}
\begin{equation}\label{63}
\begin{array}{l}
f_2=\displaystyle\frac{3Qx^4-8d_1x^3+4\left(2A-B-Qh\right)x^2+4\left(2p_1-Qh^2-2Bh\right)}{2x\sqrt{P_8\left(x\right)}},\\ \\
f_0=\displaystyle\frac{3Qx^4+8d_1x^3+4\left(2A-B-Qh\right)x^2+4\left(2p_1-Qh^2-2Bh\right)}{2x\sqrt{P_8\left(x\right)}}.
\end{array}
\end{equation}

It is well known from the general theory of ordinary differential equations (see~\cite{ZaitsevPolyanin}), that if the Riccati equation has the form~\eqref{62}, then the substitution of the form
\begin{equation*}
u\left(x\right)=\exp\left(-\int f_2\left(x\right)w\left(x\right)dx\right)
\end{equation*}
reduces the Riccati equation to the second-order linear differential equation
\begin{equation}\label{64}
\frac{d^2u}{dx^2}+a\left(x\right)\frac{du}{dx}+b\left(x\right)u=0,\quad a\left(x\right)=-\frac{1}{f_2}\frac{df_2}{dx},\quad
b\left(x\right)=f_0f_2.
\end{equation}

Taking into the account that functions $f_2$ and $f_0$ has the form~\eqref{63}, we can write the coefficients $a\left(x\right)$ and $b\left(x\right)$ in explicit form. Indeed
\begin{equation*}
a\left(x\right)=\frac{P_{12}\left(x\right)}{xP_8\left(x\right)P_4\left(x\right)},\quad b\left(x\right)=\frac{P_4\left(x\right)Q_4\left(x\right)}{4x^2P_8\left(x\right)},
\end{equation*}

\begin{equation*}
\begin{array}{l}
P_4\left(x\right)=3Qx^4-8d_1x^3+4\left(2A-B-Qh\right)x^2-4\left(Qh^2+2Bh-2p_1\right),\\ \\
Q_4\left(x\right)=16d_1x^3+P_4\left(x\right),
\end{array}
\end{equation*}

\begin{equation*}
\begin{array}{l}
P_{12}\left(x\right)=-3Q^3x^{12}+16d_1Q^2x^{11}-12Q^2\left(2A-B-Qh\right)x^{10}-64d_1\left(BQ+Q^2h-2\right)x^9+\\ \\
+64\left(BQ+Q^2h-2\right)\left(2A-B-Qh\right)x^8+44Q^2\left(Qh^2+2Bh-2p_1\right)x^8+48Q\left(B+Qh\right)^2x^8-\\ \\
-192Qhx^8+32\left(8-9BQ-9Q^2h-2AQ\right)\left(Qh^2+2Bh-2p_1\right)x^6+\\ \\
+64\left(4h-\left(B+Qh\right)^2\right)\left(2A-B-Qh\right)x^6+384Q\left(h^2-1\right)x^6+\\ \\
+256d_1\left(B+Qh\right)\left(Qh^2+2Bh-2p_1\right)x^5-512d_1\left(h^2-1\right)x^5+\\ \\
+240Q\left(Qh^2+2Bh-2p_1\right)^2x^4-192\left(4h-\left(B+Qh\right)^2\right)\left(Qh^2+2Bh-2p_1\right)x^4-\\ \\
-256d_1\left(Qh^2+2Bh-2p_1\right)^2x^3+64\left(2A-5B-5Qh\right)\left(Qh^2+2Bh-2p_1\right)^2x^2+\\ \\
+512\left(h^2-1\right)\left(Qh^2+2Bh-2p_1\right)x^2+64\left(Qh^2+2Bh-2p_1\right)^3.
\end{array}
\end{equation*}

It is easy to see, that the coefficients $a\left(x\right)$ and $b\left(x\right)$ of the linear second order differential equations~\eqref{64} are rational functions of the independent variable $x$. Therefore, we have the following result.

\begin{thm}\label{Theorem7}
The solution of the problem of motion of a heavy gyrostat with a fixed point under the action of gyroscopic forces in the Hess -- Sretensky case is reduced to solving the second order linear homogeneous differential equation~\eqref{64} with rational coefficients. $\Box$
\end{thm}

Indeed, if the general solution of the second order linear differential equation~\eqref{64} is found in the explicit form 
\begin{equation*}
u\left(x\right)=C_1u_1\left(x\right)+C_2u_2\left(x\right),
\end{equation*}
then, using this solution, we can obtain the expression for $\varphi=\varphi\left(x\right)$, using which, according to~\eqref{59}, we can find the explicit expressions for $y=y\left(x\right)$ and $z=z\left(x\right)$ . Now, from the system of equations
\begin{equation*}
\begin{array}{l}
y\left(x\right)\nu_2+z\left(x\right)\nu_3=k_1+B\left(\displaystyle\frac{x^2}{2}-h\right)-\displaystyle\frac{Q}{2}\left(\displaystyle\frac{x^2}{2}-h\right)^2, \\ \\
y\left(x\right)\nu_3-z\left(x\right)\nu_2=-\displaystyle\frac{\sqrt{P_8\left(x\right)}}{8},
\end{array}
\end{equation*}
we obtain the explicit expressions for $\nu_2=\nu_2(x)$ and $\nu_3=\nu_3(x)$. The expression for $\nu_1=\nu_1(x)$ can be found out from the first integral~\eqref{55}
\begin{equation*}
\nu_1=h-\displaystyle\frac{x^2}{2},
\end{equation*}
and the function $x=x\left(\tau\right)$ can be obtained from the first equation of the system~\eqref{60}.

Thus, the solution of the problem of motion of a heavy gyrostat with a fixed point under the action of gyroscopic forces in the integrable Hess -- Sretensky case is reduced to solving the second-order linear differential equation~\eqref{64} with rational coefficients. Therefore, we can study the problem of existence of liouvillian solutions of the linear differential equation~\eqref{64}. To solve this problem we shall apply the Kovacic algorithm to the linear differential equation~\eqref{64}. The results of application of the algorithm to this differential equation are discussed in the next Sections.

\subsection{Application of the Kovacic algorithm to the differential equation~\eqref{64}. General case.}

So, the differential equation being investigated has the form~\eqref{64}. In this equation we make a substitution~\eqref{29} and reduce it to the form~\eqref{30}:
\begin{equation}\label{65}
\frac{d^2y}{dx^2}=R\left(x\right)y.
\end{equation}

Here the function $R\left(x\right)$ takes the form:
\begin{equation*}
R\left(x\right)=\frac{U\left(x\right)}{V\left(x\right)},\quad V\left(x\right)=\left(P_8\left(x\right)\right)^2\left(P_4\left(x\right)\right)^2.
\end{equation*}

Function $U\left(x\right)$ has a very complicated form and we do not write it here explicitly. Further we shall assume, that all the roots of the function $V\left(\right)$ are distinct. This assumption are valid for almost all values of parameters of the problem. Thus, it is easy to see, that the function $R\left(x\right)$ has 12 finite poles of the second order. Let us denote the roots of the polynomial $P_8\left(x\right)$ by $x_1$, $x_2$, $x_3$, $x_4$, $x_5$, $x_6$, $x_7$, $x_8$. Note that this polynomial contains only the terms of even degree, therefore its roots satisfy the conditions:
\begin{equation*}
x_2=-x_1,\quad x_4=-x_3,\quad x_6=-x_5,\quad x_8=-x_7. 
\end{equation*}

Let us denote the roots of the polynomial $P_8\left(x\right)$ by $x_9$, $x_{10}$, $x_{11}$, $x_{12}$. Now let us consider the partial fraction expansion of the function $R\left(x\right)$. It has the form:
\begin{equation*}
R\left(x\right)=-\frac{3}{16}\sum\limits_{i=1}^8\frac{1}{\left(x-x_i\right)^2}+\sum\limits_{i=1}^{12}\frac{\gamma_i\left(x_i\right)}
{x-x_i}+\frac{3}{4}\sum\limits_{i=9}^{12}\frac{1}{\left(x-x_i\right)^2}.
\end{equation*}

The coefficients $\gamma_i\left(x_i\right)$, $i=1, 2,\ldots 9$ have a very complicated form and we do not write them explicitly here. It is possible to note the following properties of the partial fraction expansion of the function $R\left(x\right)$.
\begin{enumerate}
\item The coefficients $b_1,\ldots, b_8$ of $\displaystyle\frac{b_i}{\left(x-x_i\right)^2}$, $i=1,\ldots, 8$ are all equal
\begin{equation*}
b_i=-\frac{3}{16}, \quad i=1,\ldots, 8.
\end{equation*}
\item The coefficients $b_9$, $b_{10}$, $b_{11}$, $b_{12}$ of $\displaystyle\frac{b_i}{\left(x-x_i\right)^2}$, $i=9, 10, 11, 12$ are all equal
\begin{equation*}
b_i=\frac{3}{4}, \quad i=9, 10, 11, 12.
\end{equation*}
\item The Laurent expansion of $R\left(x\right)$ at $x=\infty$ has the form:
\begin{equation*}
R\left(x\right)=\frac{2}{x^2}+O\left(\frac{1}{x^3}\right).
\end{equation*}
\end{enumerate}

Thus, the necessary conditions of existence of liouvillian solutions of the linear differential equation~\eqref{65} are valid for any type of solutions. First we search for a solution of equation~\eqref{65} of type 1. We applied the Kovacic algorithm for the searching of this type of solution as it was described in Section 5.

\underline{{\bf Step 1.}} Let us calculate the functions $\left[\sqrt{R}\right]_c$ and constants $\alpha_c^{\pm}$. We have the following results
\begin{equation*}
\begin{array}{l}
\left[\sqrt{R}\right]_{x_i}=0,\quad \alpha_{x_i}^{+}=\displaystyle\frac{3}{4},\quad \alpha_{x_i}^{-}=\displaystyle\frac{1}{4},\quad i=1, 2, \ldots, 8;\\ \\
\left[\sqrt{R}\right]_{x_i}=0,\quad \alpha_{x_i}^{+}=\displaystyle\frac{3}{2},\quad \alpha_{x_i}^{-}=-\displaystyle\frac{1}{2},\quad i=9, 10, 11, 12;\\ \\
\left[\sqrt{R}\right]_{\infty}=0,\quad \alpha_{\infty}^{+}=2,\quad \alpha_{\infty}^{-}=-1.
\end{array}
\end{equation*}

\underline{{\bf Step 2.}} Since the number $\rho$ of finite poles of the function $R\left(x\right)$ is equal to 12, we have $2^{\rho+1}=2^{13}=8192$ tuples of signs
\begin{equation*}
s=\left(s\left(\infty\right), s\left(x_1\right), s\left(x_2\right), \ldots, s\left(x_{11}\right), s\left(x_{12}\right)\right).
\end{equation*}

Foe each tuple we evaluate the constant $d$ by the formula~\eqref{31}. According to the algorithm, $d$ must be nonnegative integer. Further, we shall analyze all possible tuples of signs $s$ and the corresponding values $\alpha$. First of all, let us consider the tuple
\begin{equation*}
s_1=\left(+, -, -, -, -, -, -, -, -, -, -, -, -\right),
\end{equation*}
for which we have $d=2$.

\underline{{\bf Step 3.}} For the tuple $s_1$ we calculate the function $\theta$ by the formula~\eqref{32}. In explicit form this function can be written as follows:
\begin{equation*}
\theta=-\frac{2xP_6\left(x\right)}{P_8\left(x\right)}-\frac{2x\left(3Qx^2-6d_1x+2\left(2A-B-Qh\right)\right)}{P_4\left(x\right)},
\end{equation*}
\begin{equation*}
\begin{array}{l}
P_6\left(x\right)=Q^2x^6+6\left(2-BQ-Q^2h\right)x^4+4\left(3Q^2h^2-8h+2B^2+6BQh-2p_1Q\right)x^2+\\ \\
+16\left(h^2-1\right)-8\left(B+Qh\right)\left(Qh^2+2Bh-2p_1\right).
\end{array}
\end{equation*}

The polynomial $P$ of degree $d=2$
\begin{equation*}
P=x^2+K_1x+K_2
\end{equation*}
should identically satisfy the differential equation~\eqref{33}. After substitution of $P$, $\theta$ and $R\left(x\right)$ to the equation~\eqref{33}, we obtain in the left hand side of~\eqref{33} the rational expression. The numerator of this expression has a form of the 11th degree polynomial:
\begin{equation*}
\Pi_{11}=-6Q^2\left(2d_1+QK_1\right)x^{11}+\cdots
\end{equation*}

Let us set
\begin{equation*}
K_1=-\frac{2d_1}{Q}.
\end{equation*}

Then the numerator of the rational expression in the left hand side of~\eqref{33} takes the form:
\begin{equation*}
\Pi_{10}=-6Q\left(2AQ+Q^2K_2+2BQ+2Q^2h-4d_1^2-6\right)x^{10}+\cdots
\end{equation*}

To turn the leading coefficient of this polynomial to zero, we set
\begin{equation*}
K_2=\frac{2\left(2d_1^2+3-Q^2h-BQ-AQ\right)}{Q^2}.
\end{equation*}

Then the numerator of the rational expression in the left hand side of~\eqref{33} takes the form:
\begin{equation*}
\Pi_9=144d_1\left(AQ-1-d_1^2\right)x^9+\cdots
\end{equation*}

To turn the leading coefficient of this polynomial to zero, we set
\begin{equation*}
A=\frac{1+d_1^2}{Q}.
\end{equation*}

Then the numerator of the rational expression in the left hand side of~\eqref{33} takes the form:
\begin{equation*}
\Pi_8=\frac{24}{Q}\left(10-10BQ-4Q^2h+4d_1^2-2hD_1^2Q^2-2Bd_1^2Q+d_1^4+2p_1Q^3+B^2Q^2\right)x^8+\cdots
\end{equation*}

To turn the leading coefficient of this polynomial to zero, we set
\begin{equation*}
p_1=\frac{2Bd_1^2Q+2hd_1^2Q^2-4d_1^2-B^2Q^2-d_1^4-10+10BQ+4Q^2h}{2Q^3}.
\end{equation*}

Then the numerator of the rational expression in the left hand side of~\eqref{33} takes the form:
\begin{equation*}
\Pi_7=\frac{480d_1}{Q^2}\left(BQ-1\right)x^7+\cdots
\end{equation*}

To turn the leading coefficient of this polynomial to zero, we set
\begin{equation*}
B=\frac{1}{Q}.
\end{equation*}

Then the numerator of the rational expression in the left hand side of~\eqref{33} takes the form:
\begin{equation*}
\Pi_6=\frac{432}{Q^3}\left(1+Q^4+2d_1^2+d_1^4-2Q^2h-2hd_1^2Q^2\right)x^6+\cdots
\end{equation*}

If we set
\begin{equation*}
h=\frac{Q^4+\left(d_1^2+1\right)^2}{2Q^2\left(d_1^2+1\right)}
\end{equation*}
then then this polynomial $\Pi_6$ becomes zero. Therefore we can state the following Theorem based on the verification of the tuple $s_1$.

\begin{thm}\label{Theorem8}
Second order linear differential equation~\eqref{64} has liouvillian solutions of type 1 if the following conditions are valid:
\begin{equation}\label{66}
A=\frac{d_1^2+1}{Q},\quad B=\frac{1}{Q},\quad h=\frac{Q^4+\left(d_1^2+1\right)^2}{2Q^2\left(d_1^2+1\right)},\quad
p_1=\frac{Q^4+\left(d_1^2+1\right)^2}{2Q^3\left(d_1^2+1\right)}+\frac{Q}{2}.\quad \Box
\end{equation}
\end{thm}

Under conditions~\eqref{66} the general solution of the second order linear differential equation~\eqref{64} takes the form
\begin{equation*}
u\left(x\right)=C_1u_1\left(x\right)+C_2u_1\left(x\right)\int\frac{f_2dx}{u_1^2\left(x\right)},\quad
u_1\left(x\right)=\frac{Q^4-\left(d_1^2+1\right)\left(\left(Qx-d_1\right)^2+1\right)}{\sqrt{x}}.
\end{equation*} 

Similarly we can check the other tuples for which the constant $d$, calculating according to~\eqref{31}, is a nonnegative integer. Checking all of these tuples gives us the same conditions as checking of the tuple $s_1$. Finally we can conclude, that conditions~\eqref{66} are necessary and sufficient conditions of existence of liouvillian solutions of type 1 of the second order linear differential equation~\eqref{64}.

Now let us study the problem of existence of liouvillian solutions of type 2 for the differential equation~\eqref{64}. The necessary conditions for the existence of such a solution are fulfilled (see Theorem~\ref{Theorem3}). We apply step by step the Kovacic algorithm as was described in Section 5.

\underline{{\bf Step 1.}} According to the Kovacic algorithm let us find the sets $E_c$ and $E_{\infty}$ for every pole of the function $R\left(x\right)$. For the finite poles $x=x_i$, $i=1,\ldots, 8$, which are roots of the polynomial $P_8\left(x\right)$, these sets $E_{x_i}$ have the form:
\begin{equation*}
E_{x_i}=\{1, 2, 3\}, \quad i=1,\ldots, 8.
\end{equation*}

For the finite poles $x=x_i$, $i=9, 10, 11, 12$, which are roots of the polynomial $P_4\left(x\right)$, these sets $E_{x_i}$ have the form:
\begin{equation*}
E_{x_i}=\{-2, 2, 6\}, \quad i=9, 10, 11, 12.
\end{equation*}

The set $E_{\infty}$ has the form
\begin{equation*}
E_{\infty}=\{-4, 2, 8\}.
\end{equation*}

\underline{{\bf Step 2.}} Now we consider all possible tuples
\begin{equation*}
s=\left(e_{\infty}, e_{x_1}, e_{x_2}, e_{x_3}, e_{x_4}, e_{x_5}, e_{x_6}, e_{x_7}, e_{x_8}, e_{x_9}, e_{x_{10}}, e_{x_{11}}, e_{x_{12}}\right)
\end{equation*}
of elements of the sets $E_{\infty}$, $E_{x_i}$, $i=1,\ldots, 12$; at least of of the elements in each tuple must be odd. Using~\eqref{34}, for each tuple $s$ we get
\begin{equation*}
d=\frac{1}{2}\left(e_{\infty}-\sum\limits_{i=1}^{12} e_{x_i}\right).
\end{equation*}

According to the algorithm, $d$ must be nonnegative integer. Note that the minimal value of the sum of the elements of sets corresponding to finite poles is zero. Therefore the maximal value of $d$, calculated according to~\eqref{34}, equals $d=4$. The value $d=4$ corresponds to the set $s_1$, in which the elements $e_{\infty} $ and $e_{x_i}$, $i=1, 2, \ldots 12$ are equal
\begin{equation*}
s_1=\left(8, 1, 1, 1, 1, 1, 1, 1, 1, -2, -2, -2, -2\right).
\end{equation*}

Let us check this tuple.

\underline{{\bf Step 3.}} According to the algorithm, let us find the function $\theta$ by the formula~\eqref{35}. Since the elements at all finite poles $x=x_i$, $i=1, 2, \ldots, 8$ are the same, and the elements at all poles $x=x_i$, $i=9, 10, 11, 12$ are the same, then we can write the function $\theta$ in explicit form. For the set $s_1$ this function has the form
\begin{equation*}
\theta=-\frac{4xP_6\left(x\right)}{P_8\left(x\right)}-
\frac{4x\left(3Qx^2-6d_1x+2\left(2A-B-Qh\right)\right)}{P_4\left(x\right)}.
\end{equation*}

The polynomial $P$ of degree $d=4$
\begin{equation*}
P=x^4+K_1x^3+K_2x^2+K_3x+K_4
\end{equation*}
should identically satisfy the differential equation~\eqref{36}. After substitution of the polynomial $P$ and the functions $\theta$ and $R\left(x\right)$ to the equation~\eqref{36}, we obtain in the left hand side of~\eqref{36} the rational expression. The numerator of this expression has a form of the 16th degree polynomial:
\begin{equation*}
\Pi_{16}=-90Q^3\left(QK_1+4d_1\right)x^{16}+\cdots
\end{equation*}

Let us set
\begin{equation*}
K_1=-\frac{4d_1}{Q}.
\end{equation*}

Then the numerator of the rational expression in the left hand side of~\eqref{36} takes the form:
\begin{equation*}
\Pi_{15}=-72Q^2\left(4AQ-12d_1^2+Q^2K_2-12+4BQ+4Q^2h\right)x^{15}+\cdots
\end{equation*}

Now let us set
\begin{equation*}
K_2=\frac{4\left(3d_1^2+3-AQ-BQ-Q^2h\right)}{Q^2}.
\end{equation*}

Then the numerator of the rational expression in the left hand side of~\eqref{36} takes the form:
\begin{equation*}
\Pi_{14}=2592d_1Q\left(AQ-1-d_1^2\right)x^{14}+\cdots
\end{equation*}

To turn the leading coefficient of this polynomial to zero we set 
\begin{equation*}
A=\frac{d_1^2+1}{Q}.
\end{equation*}

As a result the numerator of the rational expression in the left hand side of~\eqref{36} takes the form:
\begin{equation*}
\Pi_{13}=u_{13}x^{13}+u_{12}x^{12}+\cdots,
\end{equation*}

\begin{equation*}
\begin{array}{l}
u_{13}=72Q^4K_4+216Q^3d_1K_3+1728\left(d_1^2+1\right)^2-1728Q\left(d_1^2+1\right)\left(B+Qh\right)-\\ \\
-288Q^2\left(B+Qh\right)^2+288Q^2\left(B^2+6h+2Qp_1\right),\\ \\
u_{12}=\displaystyle\frac{24d_1}{Q}\Bigl(-11Q^4K_4-48Q^3d_1K_3+44Q^3\left(Qh^2+2Bh-2p_1\right)-\\ \\
-384\left(d_1^2+1\right)^2+384Q\left(B+Bd_1^2+Qhd_1^2\right)\Bigr).
\end{array}
\end{equation*}

Solving the system of two equations
\begin{equation*}
u_{13}=0,\quad u_{12}=0
\end{equation*}
with respect to unknown coefficients $K_3$ and $K_4$ of the polynomial $P$, we obtain that these coefficients are equal
\begin{equation*}
K_3=\displaystyle\frac{8\left(BQ\left(d_1^2+1\right)+Q^2hd_1^2-\left(d_1^2+1\right)^2\right)}{d_1Q^3},\quad
K_4=\displaystyle\frac{4\left(Qh^2+2Bh-2p_1\right)}{Q}.
\end{equation*} 

Substituting the expressions for $K_3$ and $K_4$ to the polynomial $\Pi_{13}$ we obtain that this polynomial becomes zero. Therefore we can state the following Theorem based on the verification of the tuple $s_1$.

\begin{thm}\label{Theorem9}
Let $Q\ne 0$ (that means that gyroscopic forces are present) and $d_1\ne 0$ (the mass distribution of the body does not correspond to the Lagrange case). Then the second-order linear differential equation~\eqref{64} admits a general solution expressed in terms of Liouvillian functions under the condition
\begin{equation}\label{67}
A=\frac{\left(d_1^2+1\right)}{Q}. \quad \Box
\end{equation}
\end{thm}

Indeed, under the condition~\eqref{67} the general solution of the linear differential equation~\eqref{64} takes the form:
\begin{equation*}
u\left(x\right)=\sqrt{\frac{G_4\left(x\right)}{x}}\left(C_1\cos\Phi\left(x\right)+C_2\sin\Phi\left(x\right)\right),\quad
\Phi\left(x\right)=8\sqrt{D}\int\frac{xf_2^*dx}{G_4\left(x\right)},
\end{equation*}
\begin{equation*}
\begin{array}{l}
G_4\left(x\right)=d_1Q^3x^4-4d_1^2Q^2x^3-4d_1Q\left(Q^2h+BQ-2-2d_1^2\right)x^2+\\ \\
+8\left(d_1^2Q^2h+\left(d_1^2+1\right)BQ-\left(d_1^2+1\right)^2\right)x+4d_1Q^2\left(Q^2h+2Bh-2p_1\right),
\end{array}
\end{equation*}
\begin{equation*}
D=\left(d_1^2+1\right)^2\left(BQ-d_1^2-1\right)^2+2d_1^2Q^2\left(d_1^2+1\right)\left(p_1Q-d_1^2h-h\right)-Q^4d_1^2,\quad
f_2^*=\left.f_2\right|_{\eqref{67}}.
\end{equation*}

Thus, the condition under which we can obtain the explicit form of the general solution of the second-order linear differential equation~\eqref{64} has the form~\eqref{67}.

\subsection{Motion of a Gyrostat Under the Action of Only Gyroscopic Forces.}

Let us consider the system~\eqref{52} and and assume that the motion is occurring only under the action of gyroscopic forces $\left(\Gamma=0\right)$. In this case the system of equations~\eqref{52} takes the form:
\begin{equation}\label{68}
\begin{array}{l}
\dot{\tilde{L}}_2=b\tilde{L}_2L_3+\left(F-Gc\right)L_3-\Lambda c\nu_1L_3,\quad \dot{L}_3=-b{\tilde L}_2^2-\left(F-Gc\right)
{\tilde L}_2+\Lambda c{\tilde L}_2\nu_1,\\ \\
\dot{\nu}_1=cL_3\nu_2-c{\tilde L}_2\nu_3,\quad \dot{\nu}_2=-cL_3\nu_1+b{\tilde L}_2\nu_3+F\nu_3,\quad
\dot{\nu}_3=c{\tilde L}_2\nu_1-b{\tilde L}_2\nu_2-F\nu_2.
\end{array}
\end{equation}

The system of equations~\eqref{68} admits the following first integrals
\begin{equation}\label{69}
\begin{array}{l}
{\tilde L}_2^2+L_3^2=E;\quad {\tilde L}_2\nu_2+L_3\nu_3+G\nu_1+\displaystyle\frac{\Lambda}{2}\nu_1^2=k;\quad
\nu_1^2+\nu_2^2+\nu_3^2=1.
\end{array}
\end{equation}

Now we shall rewrite the system~\eqref{68} and the first integrals~\eqref{69} in the dimensionless form. Since the gravity is absent, we reduce the system~\eqref{68} to the dimensionless form by the different way then the system~\eqref{52}. In particular, we introduce the dimensionless components of angular momentum $y$ and $z$ using formulas
\begin{equation*}
{\bar L}_2=y\Lambda,\quad L_3=z\Lambda.
\end{equation*}

We also introduce the dimensionless time $\tau$ using formula $\tau=\Lambda ct$ and the dimensionless parameters and constants of first integrals:
\begin{equation*}
A=\displaystyle\frac{F}{c\Lambda},\quad B=\frac{G}{\Lambda},\quad h^2=\frac{E}{\Lambda^2},\quad p_1=\frac{k}{\Lambda}.
\end{equation*}

Therefore, we can rewrite the system of equations~\eqref{68} in dimensionless form:
\begin{equation}\label{70}
\begin{array}{l}
\displaystyle\frac{dy}{d\tau}=d_1yz+\left(A-B\right)z-\nu_1z,\quad \displaystyle\frac{dz}{d\tau}=-d_1y^2-\left(A-B\right)y+\nu_1y,\\ \\
\displaystyle\frac{d{\nu}_1}{d\tau}=z\nu_2-y\nu_3,\quad \displaystyle\frac{d{\nu}_2}{d\tau}=d_1y\nu_3-z\nu_1+A\nu_3,\quad
\displaystyle\frac{d{\nu}_3}{d\tau}=-d_1y\nu_2+y\nu_1-A\nu_2,\\ \\
\end{array}
\end{equation}

The system~\eqref{70} possesses three first integrals
\begin{equation}\label{71}
y^2+z^2=h^2,\quad y\nu_2+z\nu_3+B\nu_1+\frac{\nu_1^2}{2}=p_1,\quad \nu_1^2+\nu_2^2+\nu_3^2=1.
\end{equation}

From the system~\eqref{70} using the first integrals~\eqref{71}, we obtain the second-order linear differential equation, to finding the general solution of which the solution of the problem is reduced. Let us rewrite the identity~\eqref{56} as follows:
\begin{equation*}
h^2\left(1-\nu_1^2\right)=\left(p_1-B\nu_1-\frac{\nu_1^2}{2}\right)^2+\left(\frac{d\nu_1}{d\tau}\right)^2.
\end{equation*}

Therefore, we have
\begin{equation*}
\left(\frac{d\nu_1}{d\tau}\right)^2=\frac{4\left(h^2-p_1^2\right)+8Bp_1\nu_1+4\left(p_1-h^2-B^2\right)\nu_1^2-4B\nu_1^3-\nu_1^4}{4}.
\end{equation*}

Taking the root of both parts of the equation, we get:
\begin{equation*}
\frac{d\nu_1}{d\tau}=\pm\frac{\sqrt{4\left(h^2-p_1^2\right)+8Bp_1\nu_1+4\left(p_1-h^2-B^2\right)\nu_1^2-4B\nu_1^3-\nu_1^4}}{2}.
\end{equation*}

We shall take
\begin{equation}\label{72}
\frac{d\nu_1}{d\tau}=-\frac{\sqrt{4\left(h^2-p_1^2\right)+8Bp_1\nu_1+4\left(p_1-h^2-B^2\right)\nu_1^2-4B\nu_1^3-\nu_1^4}}{2}
\end{equation}
(we can choose any of the sign before the square root in~\eqref{72}). Let us introduce the new variable $\varphi$ according to formulas
\begin{equation*}
y=h\cos\varphi,\quad z=h\sin\varphi
\end{equation*}
and assume that $\varphi=\varphi\left(\nu_1\right)$. Therefore
\begin{equation*}
\frac{dz}{d\tau}=h\cos\varphi\frac{d\varphi}{d\tau}=h\cos\varphi\frac{d\varphi}{d\nu_1}\frac{d\nu_1}{d\tau}.
\end{equation*}

Thus, we obtain the following equation for $\varphi$ from the system of equations~\eqref{70}:
\begin{equation}\label{73}
\begin{array}{l}
\displaystyle\frac{d\varphi}{d\nu_1}=\displaystyle\frac{2d_1h\cos\varphi}{\sqrt{4\left(h^2-p_1^2\right)+8Bp_1\nu_1+4
\left(p_1-h^2-B^2\right)\nu_1^2-4B\nu_1^3-\nu_1^4}}+\\ \\
+\displaystyle\frac{2\left(A-B-\nu_1\right)}{\sqrt{4\left(h^2-p_1^2\right)+8Bp_1\nu_1+4
\left(p_1-h^2-B^2\right)\nu_1^2-4B\nu_1^3-\nu_1^4}}.
\end{array}
\end{equation}

We shall further denote $\nu_1$ as $x$. Using the change of variable
\begin{equation*}
w=\tan\frac{\varphi}{2},
\end{equation*}
the equation~\eqref{73} is reduced to the Riccati equation:
\begin{equation}\label{74}
\frac{dw}{dx}=f_2w^2+f_0,
\end{equation}
\begin{equation*}
\begin{array}{l}
f_2=\displaystyle\frac{A-d_1h-B-x}{\sqrt{4\left(h^2-p_1^2\right)+8Bp_1x+4
\left(p_1-h^2-B^2\right)x^2-4Bx^3-x^4}},\\ \\
f_0=\displaystyle\frac{A+d_1h-B-x}{\sqrt{4\left(h^2-p_1^2\right)+8Bp_1x+4
\left(p_1-h^2-B^2\right)x^2-4Bx^3-x^4}}.
\end{array}
\end{equation*}

Using the substitution
\begin{equation*}
u\left(x\right)=\exp\left(-\int f_2\left(x\right)w\left(x\right)dx\right),
\end{equation*}
the Riccati equation~\eqref{74} is reduced to the second order linear differential equation:
\begin{equation}\label{75}
\frac{d^2u}{dx^2}+a\left(x\right)\frac{du}{dx}+b\left(x\right)u=0,
\end{equation}
\begin{equation*}
\begin{array}{l}
a\left(x\right)=\displaystyle\frac{S_4\left(x\right)}{\left(x+d_1h+B-A\right)\left(x^4+4Bx^3+4\left(B^2+h^2-p_1\right)x^2-8Bp_1x
+4\left(p_1^2-h^2\right)\right)},\\ \\
b\left(x\right)=-\displaystyle\frac{\left(A+d_1h-B-x\right)\left(A-d_1h-B-x\right)}{\left(x^4+4Bx^3+4\left(B^2+h^2-p_1\right)x^2-8Bp_1x
+4\left(p_1^2-h^2\right)\right)},\\ \\
S_4\left(x\right)=x^4+2\left(2B-A+d_1h\right)x^3+6B\left(B-A+d_1h\right)x^2+\\ \\
+4\left(\left(B-A+d_1h\right)\left(B^2+h^2\right)+p_1\left(A-d_1h\right)\right)x+4\left(h^2-p_1^2-Bp_1\left(B-A+d_1h\right)\right).
\end{array}
\end{equation*}

For finding the conditions of existence of liouvillian solutions of the second order linear differential equation~\eqref{75} let us apply the Kovacic algorithm to this linear differential equation. The results of this application are discussed below.

\subsection{Application of the Kovacic algorithm to the differential equation~\eqref{75}. General case.}

So, the differential equation being investigated has the form~\eqref{75}. In this equation we make a substitution~\eqref{29} and reduce it to the form~\eqref{30}:
\begin{equation}\label{76}
\frac{d^2y}{dx^2}=R\left(x\right)y.
\end{equation}

Here the function $R\left(x\right)$ takes the form:
\begin{equation*}
\begin{array}{c}
R\left(x\right)=\frac{U\left(x\right)}{V\left(x\right)},\\ \\ V\left(x\right)=\left(x+B+d_1h-A\right)^2
\left(x^4+4Bx^3+4\left(B^2+h^2-p_1\right)x^2-8Bp_1x+4\left(p_1^2-h^2\right)\right)^2.
\end{array}
\end{equation*}

Function $U\left(x\right)$ has a very complicated form and we do not write it here explicitly. Further we shall assume, that all the roots of the function $V\left(\right)$ are distinct. This assumption are valid for almost all values of parameters of the problem. Thus, it is easy to see, that the function $R\left(x\right)$ has 5 finite poles of the second order. Let us denote the roots of the polynomial 
\begin{equation*}
x^4+4Bx^3+4\left(B^2+h^2-p_1\right)x^2-8Bp_1x+4\left(p_1^2-h^2\right)
\end{equation*}
by $x_1$, $x_2$, $x_3$, $x_4$. Now let us consider the partial fraction expansion of the function $R\left(x\right)$. It has the form:
\begin{equation*}
R\left(x\right)=-\frac{3}{16}\sum\limits_{i=1}^4\frac{1}{\left(x-x_i\right)^2}+\sum\limits_{i=1}^{4}\frac{\gamma_i\left(x_i\right)}
{x-x_i}+\frac{\gamma}{x+d_1h+B-A}+\frac{3}{4\left(x+d_1h+B-A\right)^2}.
\end{equation*}

The coefficients $\gamma$ and $\gamma_i\left(x_i\right)$, $i=1, 2,\ldots 4$ have a very complicated form and we do not write them explicitly here. Let us denote the value $A-B-d_1h$ by $x_0$. It is possible to note the following properties of the partial fraction expansion of the function $R\left(x\right)$.
\begin{enumerate}
\item The coefficients $b_1,\ldots, b_4$ of $\displaystyle\frac{b_i}{\left(x-x_i\right)^2}$, $i=1,\ldots, 4$ are all equal
\begin{equation*}
b_i=-\frac{3}{16}, \quad i=1,\ldots, 4.
\end{equation*}
\item The coefficient $b$ of $\displaystyle\frac{b}{\left(x-x_0\right)^2}$ equals
\begin{equation*}
b=\frac{3}{4}.
\end{equation*}
\item The Laurent expansion of $R\left(x\right)$ at $x=\infty$ has the form:
\begin{equation*}
R\left(x\right)=\frac{3}{4x^2}+O\left(\frac{1}{x^3}\right).
\end{equation*}
\end{enumerate}

Thus, the necessary conditions of existence of liouvillian solutions of the linear differential equation~\eqref{75} are valid for any type of solutions: type 1, type 2 and type 3. First we search for a solution of equation~\eqref{75} of type 1. We apply the Kovacic algorithm for the searching of this type of solution as it was described in Section 5.

\underline{{\bf Step 1.}} Let us calculate the functions $\left[\sqrt{R}\right]_c$ and constants $\alpha_c^{\pm}$. We have the following results
\begin{equation*}
\begin{array}{l}
\left[\sqrt{R}\right]_{x_i}=0,\quad \alpha_{x_i}^{+}=\displaystyle\frac{3}{4},\quad \alpha_{x_i}^{-}=\displaystyle\frac{1}{4},\quad i=1, 2, 3, 4;\\ \\
\left[\sqrt{R}\right]_{x_0}=0,\quad \alpha_{x_0}^{+}=\displaystyle\frac{3}{2},\quad \alpha_{x_0}^{-}=-\displaystyle\frac{1}{2},
\\ \\
\left[\sqrt{R}\right]_{\infty}=0,\quad \alpha_{\infty}^{+}=\displaystyle\frac{3}{2},\quad \alpha_{\infty}^{-}=-\displaystyle\frac{1}{2}.
\end{array}
\end{equation*}

\underline{{\bf Step 2.}} Since the number $\rho$ of finite poles of the function $R\left(x\right)$ equals 5, we have $64$ tuples of signs
\begin{equation*}
s=\left(s\left(\infty\right), s\left(x_1\right), s\left(x_2\right), s\left(x_3\right), s\left(x_4\right), s\left(x_0\right)\right).
\end{equation*}

Foe each tuple we evaluate the constant $d$ by the formula~\eqref{31}. According to the algorithm, $d$ must be nonnegative integer. Further, we shall analyze all possible tuples of signs $s$ and the corresponding values $\alpha$. First of all, let us consider the tuple
\begin{equation*}
s_1=\left(+, -, -, -, -, -\right),
\end{equation*}
for which we have $d=1$.

\underline{{\bf Step 3.}} For the tuple $s_1$ we calculate the function $\theta$ by the formula~\eqref{32}. In explicit form this function can be written as follows:
\begin{equation*}
\theta=-\frac{x^3+3Bx^2+2\left(B^2+h^2-p_1\right)-2Bp_1}{x^4+4Bx^3+4\left(B^2+h^2-p_1\right)x^2-8Bp_1x+4\left(p_1^2-h^2\right)}-
\frac{1}{2\left(x+d_1h+B-A\right)}.
\end{equation*}

The polynomial $P$ of degree $d=1$
\begin{equation*}
P=x+K
\end{equation*}
should identically satisfy the differential equation~\eqref{33}. After substitution of $P$, $\theta$ and $R\left(x\right)$ to the equation~\eqref{33}, we obtain in the left hand side of~\eqref{33} the rational expression. The numerator of this expression has a form of the 3rd degree polynomial:
\begin{equation*}
\Pi_3=\left(A+B+d_1h-K\right)x^3+\cdots
\end{equation*}

Let us set
\begin{equation*}
K=A+B+d_1h.
\end{equation*}

Then the numerator of the rational expression in the left hand side of~\eqref{33} takes the form:
\begin{equation*}
\Pi_2=4Ad_1hx^2+\cdots
\end{equation*}

We suppose that $h\ne 0$ and $s_1\ne 0$. To turn the leading coefficient of this polynomial to zero, we set
\begin{equation*}
A=0.
\end{equation*}

Then the numerator of the rational expression in the left hand side of~\eqref{33} takes the form:
\begin{equation*}
\begin{array}{l}
\Pi=2h\left(d_1^3h^2+2Bh+2d_1h^2-2d_1p_1-d_1B^2\right)x-\\ \\
-2d_1hB^3+d_1^4h^4-B^4+2d_1^3h^3B-4p_1^2-4d_1hBp_1-4B^2p_1+4h^2.
\end{array}
\end{equation*}

This polynomial becomes zero if the parameters of the problem satisfy the following conditions:
\begin{equation*}
p_1=\frac{d_1^3h^2-d_1B^2+2d_1h^2+2Bh}{2d_1},\quad \left(1+d_1^2\right)\left(d_1h+B\right)^2=d_1^2.
\end{equation*}

Therefore we can state the following Theorem based on the verification of the tuple $s_1$.

\begin{thm}\label{Theorem10}
Second order linear differential equation~\eqref{75} has liouvillian solutions of type 1 if the following conditions are valid:
\begin{equation}\label{77}
A=0,\quad \left(1+d_1^2\right)\left(d_1h+B\right)^2=d_1^2,\quad p_1=\frac{d_1^3h^2-d_1B^2+2d_1h^2+2Bh}{2d_1}.\quad \Box
\end{equation}
\end{thm}

Under conditions~\eqref{77} the general solution of the second order linear differential equation~\eqref{75} takes the form:
\begin{equation*}
u\left(x\right)=C_1u_1\left(x\right)+C_2u_1\left(x\right)\int\frac{f^*_2dx}{u_1^2\left(x\right)},\quad
u_1\left(x\right)=x\sqrt{d_1^2+1}+d_1,\quad f^*_2=\left. f_2\right|_{\eqref{77}}.
\end{equation*}

Similarly we can check the other tuples for which the constant $d$, calculating according to~\eqref{31}, is a nonnegative integer. Checking all of these tuples gives us the same conditions as checking of the tuple $s_1$. Finally we can conclude, that conditions~\eqref{77} are necessary and sufficient conditions of existence of liouvillian solutions of type 1 of the second order linear differential equation~\eqref{75}.

Now let us study the problem of existence of liouvillian solutions of type 2 for the differential equation~\eqref{75}. The necessary conditions for the existence of such a solution are fulfilled (see Theorem~\ref{Theorem3}). We apply step by step the Kovacic algorithm as was described in Section 5.

\underline{{\bf Step 1.}} According to the Kovacic algorithm let us find the sets $E_c$ and $E_{\infty}$ for every pole of the function $R\left(x\right)$. For the finite poles $x=x_i$, $i=1,\ldots, 4$, which are roots of the polynomial
\begin{equation*}
x^4+4Bx^3+4\left(B^2+h^2-p_1\right)x^2-8Bp_1x+4\left(p_1^2-h^2\right)
\end{equation*}
these sets $E_{x_i}$ have the form:
\begin{equation*}
E_{x_i}=\{1, 2, 3\}, \quad i=1,\ldots, 4.
\end{equation*}

For the finite pole $x=x_0$ the corresponding set $E_{x_0}$ has the form:
\begin{equation*}
E_{x_0}=\{-2, 2, 6\}.
\end{equation*}

The set $E_{\infty}$ also has the form
\begin{equation*}
E_{\infty}=\{-2, 2, 6\}.
\end{equation*}

\underline{{\bf Step 2.}} Now we consider all possible tuples
\begin{equation*}
s=\left(e_{\infty}, e_{x_1}, e_{x_2}, e_{x_3}, e_{x_4}, e_{x_0}\right)
\end{equation*}
of elements of the sets $E_{\infty}$, $E_{x_0}$, $E_{x_i}$, $i=1,\ldots, 4$; at least of of the elements in each tuple must be odd. Using~\eqref{34}, for each tuple $s$ we get
\begin{equation*}
d=\frac{1}{2}\left(e_{\infty}-e_{x_0}-\sum\limits_{i=1}^{4} e_{x_i}\right).
\end{equation*}

According to the algorithm, $d$ must be nonnegative integer. Note that the minimal value of the sum of the elements of sets corresponding to finite poles is 2. Therefore the maximal value of $d$, calculated according to~\eqref{34}, equals $d=2$. The value $d=2$ corresponds to the tuple $s_1$, in which the elements $e_{\infty}$, $e_{x_0}$ and $e_{x_i}$, $i=1, \ldots 4$ are equal
\begin{equation*}
s_1=\left(6, 1, 1, 1, 1, -2\right).
\end{equation*}

Let us check this tuple.

\underline{{\bf Step 3.}} According to the algorithm, let us construct for the tuple $s_1$ the function $\theta$ by the formula~\eqref{35}. For the tuple $s_1$ this function has the form
\begin{equation*}
\theta=\frac{2x^3+6Bx^2+4\left(B^2+h^2-p_1\right)x-4Bp_1}{x^4+4Bx^3+4\left(B^2+h^2-p_1\right)x^2-8Bp_1x+4\left(p_1^2-h^2\right)}-
\frac{1}{x+d_1h+B-A}.
\end{equation*}

The polynomial $P$ of degree $d=2$
\begin{equation*}
P=x^2+K_1x+K_2
\end{equation*}
should identically satisfy the differential equation~\eqref{36}. After substitution of the polynomial $P$ and the functions $\theta$ and $R\left(x\right)$ to the equation~\eqref{36}, we obtain in the left hand side of~\eqref{36} the rational expression. The numerator of this expression has a form of the 4th degree polynomial:
\begin{equation*}
\Pi_4=3\left(2B+2A+2d_1h-K_1\right)x^4+\cdots
\end{equation*}

Let us set
\begin{equation*}
K_1=2A+2B+2d_1h.
\end{equation*}

Then the numerator of the rational expression in the left hand side of~\eqref{36} takes the form:
\begin{equation*}
\Pi_3=16Ad_1hx^3+\cdots
\end{equation*}

Since we assume that $d_1\ne 0$ and $h\ne 0$, we set
\begin{equation*}
A=0.
\end{equation*}

Then the numerator of the rational expression in the left hand side of~\eqref{36} takes the form:
\begin{equation*}
\Pi_2=4h\left(2d_1h^2+2Bh+2Bd_1^2h-2d_1p_1+2d_1^3h^2-K_2d_1\right)\left(x+B+d_1h\right)^2
\end{equation*}
and this polynomial becomes zero if we set
\begin{equation*}
K_2=\frac{2\left(Bh+Bd_1^2h+d_1h^2+d_1^3h^2-d_1p_1\right)}{d_1}.
\end{equation*} 

Therefore we can state the following Theorem based on the verification of the tuple $s_1$.

\begin{thm}\label{Theorem11}
The second order linear differential equation~\eqref{75} has a general solution, expressed in terms of Liouvillian functions when the condition $A=0$ is satisfied, that is in the case, when the gyrostatic momentum of the rotor is collinear to the radius--vector from the fixed point to the center of mass of the gyrostat. $\Box$
\end{thm}

Indeed, under condition $A=0$ the general solution of the equation~\eqref{43} takes the form:
\begin{equation*}
u\left(x\right)=\sqrt{G_2\left(x\right)}\left(C_1\cos\Phi\left(x\right)+C_2\sin\Phi\left(x\right)\right),\quad
\Phi\left(x\right)=2h\sqrt{D}\int\frac{f_2^*dx}{G_2\left(x\right)},
\end{equation*}
\begin{equation*}
\begin{array}{l}
G_2\left(x\right)=d_1\left(x+B+d_1h\right)^2+2hB+d_1\left(h^2d_1^2+2h^2-B^2-2p_1\right),\\ \\
D=\left(B^2-d_1^2h^2\right)\left(d_1^2+1\right)^2+2d_1^2\left(d_1^2+1\right)k_1-d_1^2,\quad
f_2^*=\left.f_2\right|_{A=0}.
\end{array}
\end{equation*}

Thus, the equations of motion of a gyrostat under the action of only gyroscopic forces in the Hess --- Sretensky integrable case can be integrated in quadratures under the condition $A=0$.

\subsection*{Conclusions.}

In this paper we discussed the problem of motion of a heavy gyrostat with a fixed point under the action of gravity and gyroscopic forces in the Hess -- Sretensky case. We proved that the solution of this problem is reduced to solving the second order linear differential equation with rational coefficients. By application of the Kovacic algorithm to this linear differential equation we found the conditions, under which the general solution of the corresponding equation can be expressed in terms of Liouvillian functions. In this case the equations of motion of a gyrostat can be integrated in quadratures. In particular, we proved that if the motion of the gyrostat occurs under the action of only gyroscopic forces, the equations of motion can be integrated in quadratures if the gyrostatic momentum of the rotor is collinear to the radius--vector from the fixed point to the center of mass of the gyrostat.


\begin{thebibliography}{99}

\bibitem{Hess}
{\em Hess~W.} Ueber die Euler'schen Bewegungsgleichungen und \"uber eine neue partikul\"are L\"osung des Problems der Bewegung eines starren K\"orpers um einen festen Punkt // Mathematische Annalen. 1890. Bd.~37. Heft~2. S.~153--181.

\bibitem{Sretensky}
{\em Sretenskii~L.~N.} Some integrability cases for the equations of gyrostat motion // Doklady Akademii Nauk SSSR. 1963. Vol.~149. P.~292--294 (in Russian).

\bibitem{Golubev}
{\em Golubev V.V.} Lectures on Integration of the Equations of Motion of a Rigid Body about a Fixed Point. Israel: Israeli Program for Scientific Translations. 1960. 287~p.

\bibitem{BorisovMamaev1}
{\em Borisov A.V., Mamaev I.S.} The Hess case in rigid body dynamics // Journal of Applied Mathematics and Mechanics. 2003. Vol.~67. P.~227--235.

\bibitem{BorisovMamaev2}
{\em Borisov A.V., Mamaev I.S.} Rigid Body Dynamics. de Gruyter Studies in Mathematical Physics. V. 52.  Walter de Gruyter Gmbh. 2019. 520~p.

\bibitem{GGK}
{\em Gashenenko I.N., Gorr G.V., Kovalev A.M.} Classical Problems in the Dynamics of Rigid Body. Kiev: Naukova Dumka. 2012. 401~p. (in Russian).

\bibitem{BurovKarapetyan}
{\em Burov A.A., Karapetyan A.V.} The Motion of a Solid in a Flow of Particles // Journal of Applied Mathematics and Mechanics. 1993. Vol.~57. P.~295--299.

\bibitem{Lunev}
{\em Lunev V.V.} Integrable cases in the problem of the motion of a heavy rigid body with a fixed point in a Lorentz force field // Doklady Akademii Nauk SSSR. 1984. Vol.~275. № 4. P.~824--826 (in Russian).

\bibitem{Kozlov}
{\em Kozlov V.V.} The problem of rigid body rotation in a magnetic field // Izvestiia Akademii Nauk SSSR. Mekhanika Tverdogo Tela. 1985. № 6. P.~28--33 (in Russian).

\bibitem{Samsonov}
{\em Samsonov V.A.} On a body rotating in a magnetic field // Izvestiia Akademii Nauk SSSR. Mekhanika Tverdogo Tela. 1984. № 4. 
P.~32--34 (in Russian).

\bibitem{Kholostova}
{\em Kholostova O.V.} On the Dynamics of a Rigid Body in the Hess Case at High-Frequency Vibrations of a Suspension Point // Russian Journal of Nonlinear Dynamics. 2020. Vol.~16. No.~1. P.~59--84.

\bibitem{BBM}
{\em Bizyaev I.A., Borisov A.V., Mamaev I.S.} The Hess -- Appelrot system and its nonholonomic analogs // Proceedings of the Steklov Institute of Mathematics. 2016. Vol.~294. P.~252--275.

\bibitem{GorrMaznev}
{\em Gorr G.V., Maznev A.A.} About motion of symmetric gyrostat with a variable gyrostatic moment in two tasks of dynamics // Russian Journal of Nonlinear Dynamics. 2012. Vol.~8. No.~2. p.~369--376.

\bibitem{Kosov}
{\em Kosov A.A.} On Analogues of the Hess Case for a Gyrostat under the Action of the Moment of Gyroscopic and Circular Forces. Mechanics of Solids. 2022. Vol.~57. No.~6. P.~1848--1861.

\bibitem{Nekrasov1} %
{\em Nekrasov P.A.} On the problem of motion of a heavy rigid body about a fixed point // Mathematicheskii Sbornik. 1892. Vol.~16. No.~2. P.~508--517.

\bibitem{Nekrasov2} %
{\em Nekrasov P.A.} Recherches analytiques sur un cas de rotation d'un solide pesant autour d'un point fixe // Mathematische Annalen. 1896. Vol.~47. P.~445--530.

\bibitem{Kovacic}
{\em Kovacic J.} An algorithm for solving second order linear homogeneous differential equations // Journal of Symbolic Computation. 1986. V.~2. P.~3--43.

\bibitem{Yehia}
{\em Yehia H.M.} Rigid Body Dynamics: A Lagrangian Approach. Cham: Springer Nature Switzerland AG. 2022. 473~p.

\bibitem{Wittenburg}
{\em Wittenburg J.} Dynamics of systems of rigid bodies. Stuttgart: Teubner. 1977. 223~p. 

\bibitem{Gavrilov}
{\em Gavrilov L.} Nonintegrability of the equations of heavy gyrostat // Compositio Mathematica. 1992. Vol.~82. No.~3. P.~275--291.

\bibitem{Volterra} 
{\em Volterra V.} Sur la th\'eorie des variations des latitudes // Acta Mathematica. 1899. Vol.~22. P.~201--358.

\bibitem{Yehia1986} 
{\em Yehia H.M.} New integrable cases in the dynamics of rigid bodies // Mechanics Research Communications. 1986. Vol.~13. No.~3. P.~169--172.

\bibitem{Komarov} 
{\em Komarov I.V.} A generalization of the Kovalevskaya top // Physics Letters. 1987. Vol.~123. P.~14--15.

\bibitem{Kharlamov1}
{\em Kharlamov P.V.} Kinematic interpretation of the motion of a body with a fixed point // Journal of Applied Mathematics and Mechanics. 1964. Vol.~28. No.~3. P.~615--621.

\bibitem{Kharlamov2}
{\em Kharlamov P.V.} Lectures on the Rigid Body Dynamics. Novosibirsk: Novosibirsk University Publishing. 1965. 265~p. (in Russian).

\bibitem{BardinKuleshov1}
{\em Bardin B.S., Kuleshov A.S.} The Kovacic algorithm and its application to the problems of classical mechanics. Moscow: Moscow Aviation Institute Publishing. 2020. 260~p. (in Russian).

\bibitem{BardinKuleshov2}
{\em Bardin B.S., Kuleshov A.S.} Application of the Kovacic algorithm for the investigation of motion of a heavy rigid body with a fixed point in the Hess case // ZAMM. Zeitschrift f\"ur Angewandte Mathematik und Mechanik. 2022. Vol.~102. №~11.

\bibitem{Novikov}
{\em Novikov M.A.} On Stationary Motions of a Rigid Body under the Partial Hess Integral Existence // Mechanics of Solids. 2018. Vol.~53. No.~3. P.~262--270.

\bibitem{ZaitsevPolyanin}
{\em Polyanin A.D., Zaitsev V.F.} Handbook of Exact Solutions for Ordinary Differential Equations. Boca Raton--New York: CRC Press. 2003. 803~p.


\end{thebibliography}
\end{document}